\documentclass[aps,pre,onecolumn]{revtex4}
\usepackage{graphicx, bm, bbm,amsmath, amssymb,amsfonts, epsfig,color,comment}

\pdfoutput=1

\newcommand{\bq}{\begin{align}}
\newcommand{\eq}{\end{align}}

\begin{document}

\title{Enhancing heat transfer in a channel with unsteady flow perturbations}

\author{Silas Alben$^*$, Shivani Prabala, and Mitchell Godek}
%\thanks{}
\affiliation{Department of Mathematics, University of Michigan,
Ann Arbor, MI 48109, USA}
\email{alben@umich.edu}

\begin{abstract}
We compute unsteady perturbations that optimally increase the heat transfer (Nu) of optimal steady unidirectional channel flows, for a given average rate of power consumption Pe$^2$. The perturbations are expanded in a basis of modes, and the heat transfer enhancement corresponds to eigenvalues of the Hessian matrix of second derivatives of the Nusselt number with respect to the mode coefficients. Enhanced heat transfer, i.e. positive eigenvalues, occur in a range of temporal periods $\tau$ that scale as Pe$^{-1}$. At small to moderate $\tau$Pe values the corresponding flows are chains of eddies near the walls that move as traveling waves at the steady background flow speed. At large $\tau$Pe the flows have eddies of multiple scales ranging up to the domain size. We use an unsteady solver to simulate these flows with perturbation sizes ranging from small to large, and find increases in Nu of up to 56\% at Pe = 2$^{19}$. Large Nu can be obtained by eddies with small spatial/temporal scales and by eddies with a range of spatial scales and large temporal scales.
\end{abstract}


\maketitle
%\begin{comment}
\section{Heat transfer in a channel}

A wide variety of methods have been proposed to enhance heat transfer from solid surfaces to adjacent fluid flows \cite{webb2005enhanced}.  Passive strategies include roughening the solid surface or attaching small rigid or flexible obstacles that create vortical structures and enhance thermal mixing \cite{dipprey1963heat,gee1980forced,hart1985heat,fiebig1991heat,tsia1999measurements,promvonge2010enhanced,Castelloes2010}. Similar flow manipulations can be achieved by actively forcing the fluid (and possibly the heated solid surface) on small scales. Electric or magnetic fields, traditional fans, or piezoelectric devices may be applied \cite{webb2005enhanced,accikalin2007characterization,bergles2013current}. For example, flapping plates or synthetic jets (small cavities with oscillating walls) generate vortical structures that impinge on heated surfaces \cite{chaudhari2010heat,park2016enhancement,glezer2016enhanced,gallegos2017flags,lee2017heat,rips2020heat}. These manipulations typically increase the rate of heat transfer but also require additional power input. Passive strategies require additional power to move the flow past the obstacles, and active devices apply additional power directly to the fluid. The overall goal is to increase the average rate of heat transfer for a given average power needed to generate the flow. The simplest term in the power is the ``pumping power," the power needed to drive the flow past the heated surfaces and any additional obstacles in the flow domain \cite{han1985heat,webb2005enhanced,karwa2013performance,alben2017optimal}. At steady state, this is the time-averaged rate of viscous dissipation in the flow domain \cite{alben2017improved}. Power losses also occur internally in the devices that drive the bulk flow and apply small-scale forcings. Because this second contribution is device-specific, we focus on the average rate of viscous dissipation in the flow, which can be considered apart from the specific devices needed to generate the flow.

Here we study the problem of optimal flows for heat transfer in one of the most general and widely-studied geometries: a channel or duct with heated walls. The flows and temperature fields under the most basic boundary conditions are treated in textbooks, e.g. \cite{bergman2011fundamentals,lienhard2013heat}, where circular ducts are discussed.
\cite{shah2014laminar} is a more comprehensive reference book that includes rectangular ducts.

In previous work, we applied an optimization approach to find the 2D steady flows in a channel that maximize the Nusselt number Nu, the rate of heat transfer, for a given rate of viscous dissipation Pe$^2$ \cite{alben2017improved}. For Poiseuille flow, a similarity solution shows that Nu $\sim$ Pe$^{1/3}$ \cite{leveque1928laws}. The optimal flow improves on Poiseuille flow by increasing the flow speed in the thermal boundary layer, thereby increasing the advection of heat out of the channel. The optimum occurs when the boundary layer thicknesses of the temperature and flow velocity match, both scaling as Pe$^{-2/5}$, resulting in Nu $\sim$ Pe$^{2/5}$ \cite{alben2017improved}. All of the viscous power is expended in the boundary layer, and outside the flow is uniform with speed $\sim$ Pe$^{4/5}$. These optimal solutions were found computationally, by solving the Euler-Lagrange equations for the PDE-constrained optimization problem, with the solution consisting of the optimal flow field, the optimal temperature field, and Lagrange multipliers that enforce the advection-diffusion equation and the power constraint. The Euler-Lagrange equations are a coupled system of three unsteady nonlinear PDEs with fourth-order spatial derivatives and an integral constraint, and were solved using Newton's method. A very fine mesh is needed near the boundary to resolve the boundary layer structure, and thus the equations are difficult to solve at large Pe using Newton's method, as the Jacobian is ill-conditioned due to the fourth-order derivatives and a small mesh spacing. The optimal 2D flows were approximately unidirectional, $(u(x,y),v(x,y)) \approx (u(y), 0)$, with viscous dissipation confined to a boundary layer. To obtain better resolution of the boundary layer, a Broyden's method was developed to solve for $u(y)$ in the boundary layer region up to Pe = 2$^{19}$, and this gave the scalings above. 

Many of the heat transfer enhancement strategies above induce unsteady flow structures in the thermal boundary layer. Therefore, in this paper we search for {\it unsteady} optimal flows, and determine if they can outperform the steady optima of \cite{alben2017improved}. In the unsteady case, the Euler-Lagrange equations are essentially the same, with the addition of the time-derivative in the advection-diffusion equation and minus the time-derivative in the equation for the Lagrange multiplier (called $m$ in \cite{alben2017improved}). The equations have the same ill-conditioning issue, but the system is now much larger and more expensive to solve with the addition of the time dimension to the two space dimensions. As a first step towards finding unsteady optima in general, we search for optima within the class of small unsteady perturbations of the optimal steady unidirectional flows. This restriction greatly decreases the cost of the computations, but provides significantly improved heat transfer, as we will show. 
Although the unsteady optima are derived in the limit of small perturbations, they can also be evaluated with large perturbations, and here we find large improvements over the steady case.

\section{Model \label{Model}}

\begin{figure}
    \centering
\includegraphics[width=0.5\textwidth]{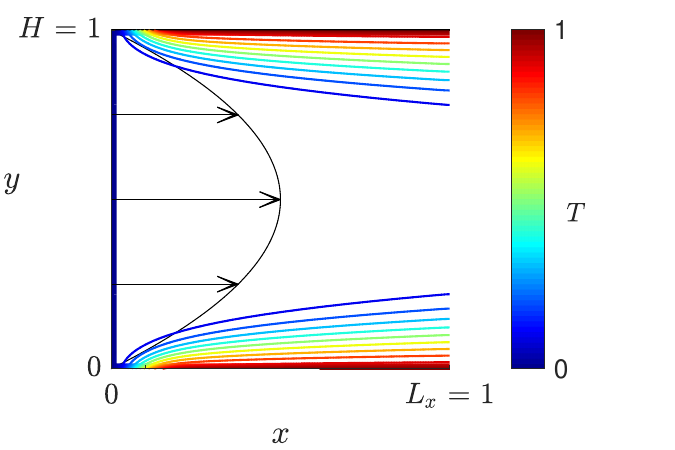}
    \caption{\footnotesize Channel geometry and temperature contours for a Poiseuille flow of the form $u(y) = 512\sqrt{3}\left(y-y^2\right)$. The temperature is colored along the top, bottom, and left (inflow) boundaries according to the boundary conditions.}
\label{fig:PoiseuilleFlowTempFigNew}
             \vspace{-.10in}
\end{figure}

We consider a 2D rectangular channel with height $H$ and horizontal length $L_x$, shown in figure \ref{fig:PoiseuilleFlowTempFigNew}. We set $H = 1$ which effectively nondimensionalizes lengths by the channel height. 
Given a flow $(u(x,y,t),v(x,y,t))$ in the channel ($0 \leq x \leq L_x$, $0 \leq y \leq 1$), we solve the advection-diffusion equation for the temperature $T$:
\begin{align}
    \partial_tT + u\partial_x T + v\partial_y T - \nabla^2T = 0. \label{AdvDiff}
\end{align}
The prefactor of $\partial_{xx}T + \partial_{yy}T$ is the thermal diffusivity $\kappa$, and is set to unity here, corresponding to nondimensionalizing velocities by $\kappa/H$. Figure \ref{fig:PoiseuilleFlowTempFigNew}
shows contours of the temperature field for the simple case of steady Poiseuille flow. This is a unidirectional flow ($(u(x,y,t),v(x,y,t)) = (u(y),0)$) in which $u$ is a quadratic function of $y$ that corresponds to a uniform horizontal pressure gradient in the channel. 

For all the flows we consider, we set the temperature boundary conditions as follows. At the channel entrance $x = 0$, we set $T = 0$. On the top and bottom walls the temperature is $T = 1-e^{-x^2/\delta^2}$ with $\delta$ a smoothing length that is set to 0.1. So beyond a small transition region with $x \sim \delta$ the wall temperature $\approx$ 1. Thus the temperature has been nondimensionalized by the difference between this limiting value and the inlet temperature. At the channel exit $x = L_x$, we use the outflow condition $\partial_x T = 0$.  

In the example of figure \ref{fig:PoiseuilleFlowTempFigNew}, the contours show that the temperature drops from 1 to 0 over a region near the boundary. In general, the faster the flow through the channel, the thinner the boundary layer, and for all the flows considered here the boundary layer will be thinner than that in figure \ref{fig:PoiseuilleFlowTempFigNew}, usually much thinner. In this case the top and bottom walls are decoupled, so $H$ is not an important length scale. As explained in \cite{alben2017improved}, one can then use $L_x$ as the characteristic length scale. Hence, in this work we set $L_x = 1$ but expect the results to apply to other $L_x$ by rescaling. We still use the full channel geometry with both walls shown in figure \ref{fig:PoiseuilleFlowTempFigNew}, in order to check whether any nontrivial interactions between the walls arise for optimal flows.

We assume that the flow is time periodic with (dimensionless) period $\tau$, and so is the temperature field.
We wish to find the incompressible flow field that maximizes the time-averaged heat transfer from the top and bottom walls of the channel,
\begin{align}
    \mbox{Nu} &= \frac{1}{\tau}\int_0^\tau \int_0^{L_x} \partial_y T \Big|_{y = 1} -\partial_y T \Big|_{y = 0} dx \, dt. \label{Nu}
\end{align}
To automatically obey incompressibility of the flow ($\partial_x u + \partial_y v = 0$), we write the flow in terms of the stream function $\psi(x,y,t)$, so $(u,v) = (\partial_y \psi, -\partial_x \psi)$.

The time-averaged power (per unit out-of-plane width) needed to maintain the flow is
the time-averaged rate of viscous energy dissipation:
\begin{align}
   \mbox{Power} = \frac{1}{\tau} \int_0^\tau \int_0^1 \int_0^{L_x}  2e_{ij}^2\, dx\, dy\, dt,
\end{align}
with the spatial integral taken over the flow domain. Here $e_{ij} = (\partial_{x_i}u_j + \partial_{x_j}u_i)/2$ is the symmetric part of the velocity gradient tensor and $e_{ij}^2$ is the sum of the squares of its entries. Thus
\begin{align}
    2 e_{ij}^2 &= 2 \partial_x u^2 + (\partial_y u + \partial_x v)^2 + 2\partial_y v^2.  
\end{align}
We have nondimensionalized the power by $\mu W\kappa^2/H^4$ where $\mu$ is the fluid viscosity and $W$ is the out-of-plane width of the flow domain.  
We now write $u$ and $v$ in terms of $\psi$ to obtain:
\begin{align}
   \mbox{Power} &= \frac{1}{\tau}\int_0^\tau \int_0^1 \int_0^{L_x} \left(\partial_{xx}\psi - \partial_{yy}\psi\right)^2 + 4\partial_{xy}\psi^2  \,dx\,dy\,dt = \frac{1}{\tau}\int_0^\tau \int_0^1 \int_0^{L_x} \left(\nabla^2\psi\right)^2  \,dx\,dy\,dt. \label{Power}
\end{align}
The second equality in (\ref{Power}) holds when the component of the flow tangent to the boundary is zero, as is the case here \cite[Art. 329]{lamb1932hydrodynamics}. We maximize Nu over the set of flows with Power = Pe$^2$, a constant. Pe stands for ``P\'{e}clet number," a measure of the dimensionless flow velocity.

In \cite{alben2017improved} we computed optimal {\it steady} flows that maximize Nu over a wide range of Pe values.
Now we compute optimal {\it unsteady} flows, assuming they are small unsteady perturbations of the optimal steady flows. This assumption greatly decreases the cost of the computations, as we will discuss, but we will study the resulting perturbations at both small and large amplitudes.

\begin{figure}
    \centering
\includegraphics[width=1\textwidth]{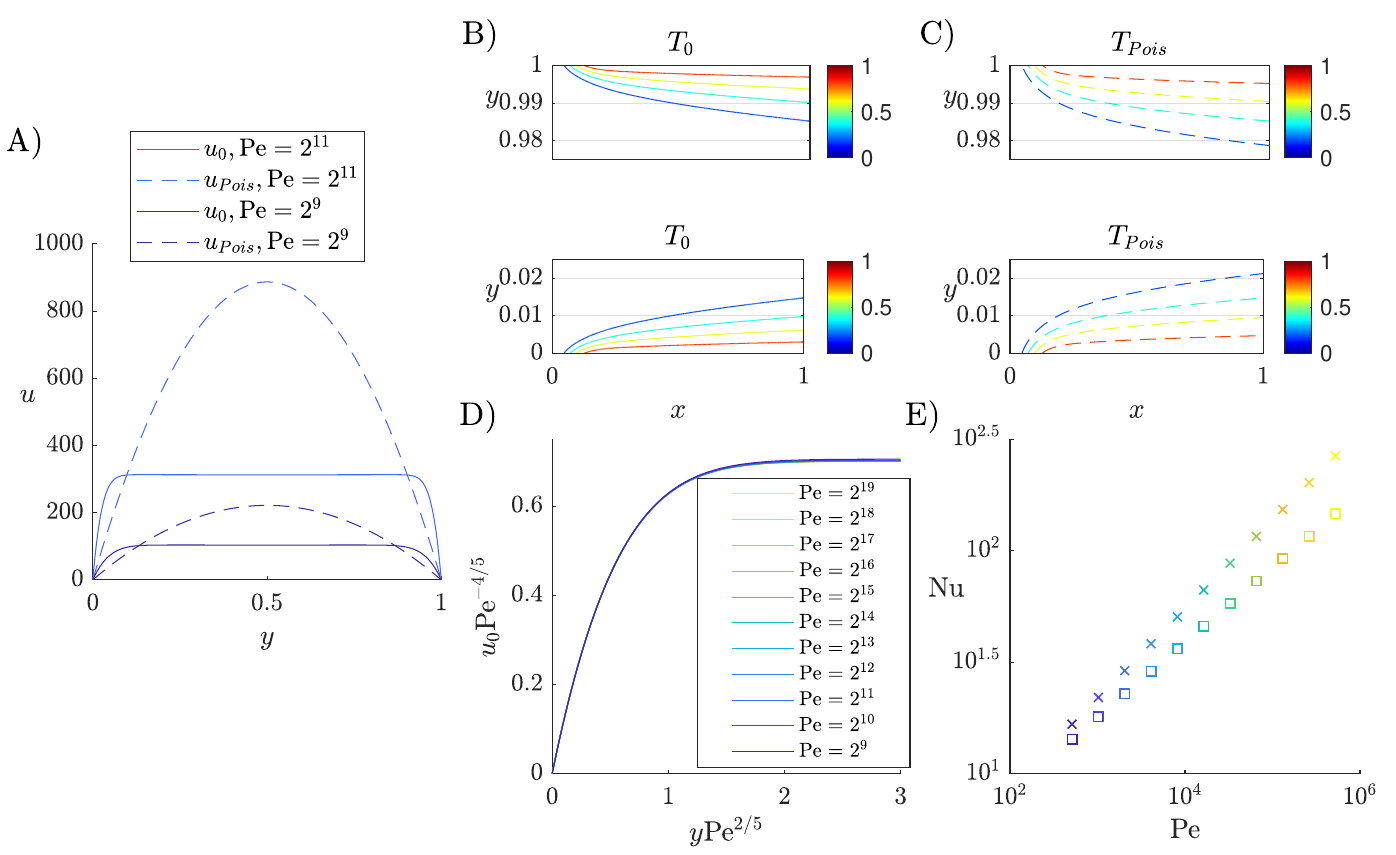}
    \caption{\footnotesize Examples of optimal unidirectional flows computed in \cite{alben2017improved}. A) Comparison of optimal unidirectional flows $u_0(y)$ and
    Poiseuille flows $u_{Pois}(y)$, at two Pe values, $2^9$ and $2^{11}$. B, C) Contours for the temperature field (at values $T = 0.2, 0.4, 0.6,$ and 0.8) near the top and bottom walls for the optimal unidirectional flow (B) and Poiseuille flow (C) at Pe = 2$^{16}$. D) Rescaled optimal unidirectional flows showing self-similar profiles. E) Nu versus Pe for the optimal unidirectional flows (crosses) and Poiseuille flow (squares) at the same Pe as in panel D.}
    \label{fig:ChannelSteadyOptFig}
             \vspace{-.10in}
\end{figure}
%%%
In \cite{alben2017improved} we solved the Euler-Lagrange equations to compute optimal 2D steady flows given as $\psi(x,y)$. These flows were nearly unidirectional, i.e. $u(x,y) 
= \partial_y\psi(x,y) \approx u(y)$.
The flow velocity (and temperature field) varied sharply within a thermal/flow boundary layer next to each channel wall and were nearly uniform outside the boundary layer, in the channel bulk. In order to obtain better resolution of the optimal flows and particularly the boundary layers, we developed a second optimization approach that explicitly assumed a unidirectional flow $u(y)$ that tends to a constant for $y = O(1)$. By solving the resulting Euler-Lagrange equations we obtained the optimal unidirectional flows $u_0(y)$ shown in figure  \ref{fig:ChannelSteadyOptFig}A and D. Panel A compares optimal unidirectional flows at Pe = $2^9$ and $2^{11}$ with the corresponding Poiseuille flows. Outside of boundary layers the unidirectional flows are uniform and hence the local power density $(\nabla^2 \psi)^2$ is zero. The power is fully concentrated in the boundary layers, allowing for a faster flow there, which convects hot fluid out of the channel faster. The total rate of convection of heat out of the channel is $\int_0^1 u_0 T \,dy$ at $x = 1$ and equals Nu in (\ref{Nu}), as can be shown by integrating the steady version of (\ref{AdvDiff}) over the channel and using the divergence theorem and flow incompressibility. The local rate of heat transfer along the heated walls, $|\partial_y T|$, is larger for the optimal flow (panel B) than for Poiseuille flow (panel C) at Pe = $2^{16}$, and this is indicated by the larger density of contour lines near the boundary in panel B. The optimal flows have a self-similar velocity profile at large Pe given by $u_0(y ; \mbox{Pe}) \approx \mbox{Pe}^{4/5}g(y \mbox{Pe}^{2/5})$ for a universal function $g$. In panel D the rescaled velocity profiles nearly coincide for Pe = 2$^9$--$2^{19}$ and the graphs approximate $g$. In \cite{alben2017improved} we explained these power law scalings using similarity solutions for the temperature profiles in uniform and linear shear flows \cite{graetz1882ueber,leveque1928laws} and optimizing the scaling exponent of Nu. The optimal flows result from matching the velocity and thermal boundary layer thicknesses. Panel D shows that the velocity boundary layer thickness $\sim$ Pe$^{-2/5}$. Since this is also the thermal boundary layer thickness, $T$ varies from 1 to 0 over this length scale, so $|\partial_y T| \sim $Pe$^{2/5}$, and Nu has the same scaling. Nu versus Pe for the optimal flows is shown by the crosses in panel E, and closely follows a Pe$^{2/5}$ scaling. The same data for Poiseuille flows is shown by the squares and follows a lower Pe$^{1/3}$ scaling \cite{,leveque1928laws}.

Now we look for optimal {\it unsteady} perturbations of the optimal steady flows, which we write in terms of the stream function as $\psi_0(y)$ instead of $u_0(y)$.
We first consider an unsteady flow of the form
\begin{align}
\psi(x,y,t) = \frac{1}{\sqrt{1+c_1\epsilon^2}}\psi_0(y) + \frac{\epsilon}{\sqrt{1+c_1\epsilon^2}}f_A(x,y)\cos(2\pi t/\tau)
+ \frac{\epsilon}{\sqrt{1+c_1\epsilon^2}}f_B(x,y)\sin(2\pi t/\tau). \label{psipert}
\end{align}
Here $0 < \epsilon \ll 1$ and the 
constant $c_1$ will be chosen so that $\psi$ has time-averaged power Pe$^2$. 
We will now compute Nu for $\psi$ by expanding both $\psi$ and $T$ in powers of $\epsilon$,
\begin{align}
    \psi(x,y,t) &= \psi_0(y) + \epsilon\psi_1(x,y,t) + \epsilon^2\psi_2(x,y,t) + \ldots \label{Psiexpn}\\
    T(x,y,t) &= T_0(x,y) + \epsilon T_1(x,y,t) + \epsilon^2 (T_{2s}(x,y) + T_{2u} (x,y,t)) + \ldots, \label{Texpn}
\end{align}
plugging both series into equation (\ref{AdvDiff}), and solving the equation at each power of $\epsilon$. Then we plug the $T$ expansion
(\ref{Texpn}) into (\ref{Nu}) to obtain a similar expansion for Nu:
\begin{align}
    \mbox{Nu} &= \mbox{Nu}_0 + \epsilon \mbox{Nu}_1 + \epsilon^2 \mbox{Nu}_2 + \ldots. \label{Nuexpn}
\end{align}
In (\ref{Texpn}) we have separated $T_2$ into its time-average $T_{2s}$ (the steady part) and the remainder $T_{2u}$ (the unsteady part); only the steady part will be needed for Nu.
At $O(\epsilon^0)$ equation (\ref{AdvDiff}) is
\begin{align}
    \partial_y\psi_0 \partial_xT_0 - \partial_x\psi_0\partial_yT_0 - \nabla^2T_0 = 0. \label{T0}
\end{align}
and thus Nu$_0$ is the Nusselt number for the optimal steady flow $\psi_0$.
At $O(\epsilon^1)$ equation (\ref{AdvDiff}) is
\begin{align}
    \partial_t T_1 + \partial_y\psi_0 \partial_xT_1 - \partial_x\psi_0\partial_yT_1 - \nabla^2T_1 = - \partial_y\psi_1 \partial_xT_0 + \partial_x\psi_1\partial_yT_0. \label{T1}
\end{align}
Since the right-hand side has components proportional to $\cos(2\pi t/\tau)$ and $\sin(2\pi t/\tau)$, the solution can be written as $T_1 = T_{1A}(x,y)\cos(2\pi t/\tau) + T_{1B}(x,y)\sin(2\pi t/\tau)$. Inserting this expression
into (\ref{T1}) gives a coupled system for $T_{1A}$ and $T_{1B}$:
\begin{align}
    \frac{2\pi}{\tau}T_{1B} + \partial_y\psi_0 \partial_xT_{1A} - \partial_x\psi_0\partial_yT_{1A} - \nabla^2T_{1A} &= -\partial_y f_A \partial_xT_0 +  \partial_x f_A\partial_yT_0. \label{T1A} \\
     -\frac{2\pi}{\tau} T_{1A} + \partial_y\psi_0 \partial_xT_{1B} - \partial_x\psi_0\partial_yT_{1B} - \nabla^2T_{1B} &= -\partial_y f_B \partial_xT_0 + \partial_x f_B\partial_yT_0.
    \label{T1B}
\end{align}

Since $T_1$ has zero time-average, Nu$_1 = 0$. We need to compute higher-order terms in Nu to determine whether the unsteady flow perturbation can increase Nu. 
At $O(\epsilon^2)$ equation (\ref{AdvDiff}) is
\begin{align}
    \partial_t T_2 + \partial_y\psi_0 \partial_xT_2 - \partial_x\psi_0\partial_yT_2 - \nabla^2T_2 = - \partial_y\psi_1 \partial_xT_1 + \partial_x\psi_1\partial_yT_1 - \partial_y\psi_2 \partial_xT_0 + \partial_x\psi_2\partial_yT_0. \label{T2}
\end{align}
Equation (\ref{psipert}) gives $\psi_2 = -c_1\psi_0/2$. Since $\psi_1$ and $T_1$ are proportional to $\cos(2\pi t/\tau)$ and $\sin(2\pi t/\tau)$, the right side of (\ref{T2}) has steady terms and terms proportional to 
$\cos(4\pi t/\tau)$ and $\sin(4\pi t/\tau)$.
The unsteady terms have zero time-average, so only the steady terms make a nonzero contribution to Nu$_2$. We compute it by first solving for $T_{2s}$:
\begin{align}
    \partial_y\psi_0 \partial_xT_{2s} - \partial_x\psi_0\partial_yT_{2s} - \nabla^2T_{2s} = - \frac{1}{2}\partial_yf_A \partial_xT_{1A} + \frac{1}{2}\partial_xf_A\partial_yT_{1A}& \nonumber\\
    - \frac{1}{2}\partial_yf_B \partial_xT_{1B} + \frac{1}{2}\partial_xf_B\partial_yT_{1B} + \frac{c_1}{2}\partial_y\psi_0 \partial_xT_0 - \frac{c_1}{2}\partial_x\psi_0\partial_yT_0&. \label{T2s}
\end{align}
and then computing
\begin{align}
    \mbox{Nu}_2 &= \int_0^{L_x} \partial_y T_{2s} \Big|_{y = 1} -\partial_y T_{2s} \Big|_{y = 0} dx. \label{Nu2}
\end{align}
The time average is dropped in (\ref{Nu2}) because the integrand is steady.
To summarize, the leading-order change in Nu due to the unsteady flow perturbation is $\epsilon^2$Nu$_2$, and to calculate Nu$_2$ we need to calculate $T_{1A}$, $T_{1B}$, and $T_{2}$. This requires solving three steady elliptic PDEs---(\ref{T1A}), (\ref{T1B}), and (\ref{T2s})---two of which are coupled. This is typically much faster than solving the unsteady advection-diffusion equation (\ref{AdvDiff}) over many time steps with a given flow $\psi$. To obtain convergence from an initial temperature field to the periodic solution, the flow would generally need to be computed over at least a few periods, with many time steps per period to resolve the solution.
An implicit time-stepping scheme would require solving an elliptic PDE similar to the three just mentioned, but for $\gtrsim$ hundreds of time steps.  
An explicit time-stepping scheme would avoid solving an elliptic PDE at each time step, but may require very small time steps for stability, particularly at large Pe. In both implicit and explicit schemes, small time steps would be needed for accuracy at large Pe, when the magnitudes of the flow and its spatial and temporal derivatives are large.

%When we discretize, the power takes the form of an inner product
%\begin{align}
%   \langle \psi_j, \psi_k \rangle_{\mbox{Power}} &=
%   \psi_j^T \mathbf{M}\psi_k, \\
%   \mathbf{M} &\equiv  
%   (\mathbf{Wt}* \mathbf{D2x})^T \mathbf{D2x} +
%   (\mathbf{Wt}* \mathbf{D2y})^T \mathbf{D2y} \nonumber\\
%   &-2
%   (\mathbf{Wt}* \mathbf{D2x})^T \mathbf{D2y} + 4
%   (\mathbf{Wt}* \mathbf{Dxy})^T \mathbf{Dxy}. \label{Mmat}
%\end{align}
%with $\mathbf{Wt}$ a diagonal matrix of weights for the double integral.
\section{Discretization}
We solve (\ref{T1A}), (\ref{T1B}), and (\ref{T2s}) by representing each component of $\psi$ and $T$ in (\ref{Psiexpn}) and  (\ref{Texpn}) on tensor product grids in $(x,y)$ space that are concentrated at the boundaries. We start with unstretched coordinates $X$ and $Y$, and define uniform grids on $[0, 1]$ in $X$ and $Y$ with spacing $1/m$ and $m$ = 384. We then map these points to $(x,y)$ space:
\begin{align}
    x = X - \eta\frac{1}{2\pi}\sin 2\pi X \quad ; \quad y = Y - \eta\frac{1}{2\pi}\sin 2\pi Y
\end{align}
with $\eta$ a stretching factor that is set to 0.997. We obtain stretched grids with points concentrated near the boundaries in $x$ and $y$, to resolve the boundary layers. The grid spacing in $x$ and $y$ ranges from about $(1-\eta)/m \approx 8\times 10^{-6}$ at the boundaries to about $(1+\eta)/m \approx 5\times 10^{-3}$ at the domain center. We discretize all spatial derivatives with second-order finite-differences using three or four grid points, one-sided at the boundaries.
Integrals such as (\ref{Nu}) and (\ref{Power}) are discretized with the trapezoidal rule.

\section{Flow modes \label{sec:Modes}}
Our overall goal is to maximize Nu over a large space of unsteady flow perturbations. The particular form of the perturbation in (\ref{psipert}) is determined by the functions $f_A$ and $f_B$ and the scalar $\tau$. To represent a wide range of flows, we expand $f_A$ and $f_B$ using products of Chebyshev polynomials in $x$ and $y$. However, we need to ensure that that the total flow (\ref{psipert}) obeys no-slip boundary conditions on the channel walls and is directed into/out of the domain on the inflow/outflow boundaries, so the temperature boundary conditions are valid. A simple way to do this is to make the unsteady velocity zero at all four sides of the domain, so the total flow reduces to the steady optimum $\psi_0$ on all four sides, and we already know $\psi_0$ obeys the desired boundary conditions.

Our first step to create a flow mode basis for $f_A$ and $f_B$ that obeys no-slip at the four sides is to define a sequence of degree-($k+4$) polynomials, $k = 0, 1, \dots$, that are linear combinations of Chebyshev polynomials up to degree $k+4$ that have zero value and first derivative at $y = 0$ and 1. With these four constraints, the polynomials must have degree at least four to be nonzero. The desired linear combinations of Chebyshev polynomials are generated using an orthogonal projection procedure described 
in \cite[App. B]{alben2023transition}, and are the $Y_k$ defined there. We use the resulting polynomials on both the $x$ and $y$ grids and denote them as $X_j(x)$ and $Y_k(y)$ with $j,k = 0, 1, \ldots, M-1$. We form all the products $X_j(x)Y_k(y)$ on the spatial grid and assemble them as $N \equiv M^2$ columns of a matrix $\mathbf{Z}$. The products have zero value and normal derivative on the boundaries, and hence satisfy the no-slip boundary conditions.

The problem is further simplified by making the flow modes orthogonal with respect to the power inner product (\ref{Power}), in discretized form.  For a discretized flow of the form  $\mathbf{Z}_n$ multiplied by either $\cos (2\pi t/\tau)$ or  $\sin (2\pi t/\tau)$, we can write the discretized power integral (\ref{Power}) as a weighted inner product $(\sqrt{\mathbf{W}}\mathbf{L}\mathbf{Z}_n)^T\sqrt{\mathbf{W}}\mathbf{L}\mathbf{Z}_n$, where $\mathbf{L}$ is the discrete Laplacian and $\mathbf{W}$ is the diagonal matrix of weights corresponding to the compound trapezoidal rule for the spatial integrals in (\ref{Power}), multiplied by 1/2 for the time integral.
We perform a QR factorization
\begin{align}
\frac{1}{\mbox{Pe}}\mathbf{\sqrt{W}L}\mathbf{Z} = \mathbf{Q}\hat{\mathbf{R}}
\end{align}
Then the columns of $\mathbf{V} \equiv \mathbf{Z}\hat{\mathbf{R}}^{-1}$ are orthogonal with respect to the power inner product and have power Pe$^2$. These are the modes that are used as a basis for $f_A$ and $f_B$ in the unsteady part of the flow given by (\ref{psipert}). 

We use the modes of $\mathbf{V}$ to write a discrete version of the flow in (\ref{psipert}):
\begin{align}
\mathbf{\Psi} = \frac{1}{\sqrt{1+\|\mathbf{a}\|^2}} \mathbf{\Psi}_0 + \sum_{j = 1}^{N} \frac{\mathrm{a}_{j}}{\sqrt{1+\|\mathbf{a}\|^2}}
\mathbf{V}_j\cos\left(\frac{2\pi t}{\tau}\right) + 
\sum_{j = 1}^{N} \frac{\mathrm{a}_{j+N}}{\sqrt{1+\|\mathbf{a}\|^2}}
\mathbf{V}_j\sin\left(\frac{2\pi t}{\tau}\right).\label{Psi}
\end{align}
Here $\mathbf{V}_j$ are columns of $\mathbf{V}$ representing different spatial modes and $\mathbf{a}$ is a vector of coefficients for the modes. Each of the terms in (\ref{Psi}) is orthogonal with respect to the power inner product. Those with different temporal functions are orthogonal due to the time integral in (\ref{Power}), and those with different $\mathbf{V}_j$ are orthogonal by the definition of $\mathbf{V}$. Hence $\mathbf{\Psi}$ has power Pe$^2$ for any real vector $\mathbf{a}$. When 
$\|\mathbf{a}\| = O(\epsilon)$, the flow (\ref{Psi}) is a discrete version of the flow (\ref{psipert}). In this case we can write a Taylor expansion for Nu like (\ref{Nuexpn}) but in powers of $\mathbf{a}$:
\begin{align}
    \mbox{Nu}(\mathbf{a}) = \mbox{Nu}(\mathbf{0}) +  \mathbf{a}^T D\mbox{Nu}\bigg|_{\mathbf{a} = \mathbf{0}}+\frac{1}{2}\mathbf{a}^T D^2\mbox{Nu}\bigg|_{\mathbf{a} = \mathbf{0}} \mathbf{a} + \ldots.
\end{align}
where $\mbox{Nu}(\mathbf{0})$ = Nu$_0$ is Nu for the steady flow and $D\mbox{Nu}$ and $D^2\mbox{Nu}$ are the gradient vector and Hessian matrix of first and second derivatives of Nu with respect to components of $\mathbf{a}$, respectively. We compute the gradient and Hessian as the limits of finite differences. The gradient entries are
\begin{align}
   D\mbox{Nu}_i \bigg|_{\mathbf{a} = \mathbf{0}} = \frac{\partial \mbox{Nu}}{\partial \mathrm{a}_i}\bigg|_{\mathbf{a} = \mathbf{0}} = \lim_{\epsilon \to 0} \frac{\mbox{Nu}(\epsilon\mathbf{e}_i)-\mbox{Nu}(\mathbf{0})}{\epsilon} = \mbox{Nu}_1(\epsilon\mathbf{e}_i)-\mbox{Nu}_1(\mathbf{0}) = 0. \label{DNu}
\end{align}
The limit of the finite difference quotient in (\ref{DNu}) is
computed by inserting the perturbation expansion of Nu (\ref{Nuexpn}) for the flows corresponding to $\mathbf{a} =\epsilon\mathbf{e}_i$ and 
$\mathbf{a} = \mathbf{0}$. As explained below (\ref{T1B}), Nu$_1$ = 0 for $\mathbf{a} =\epsilon\mathbf{e}_i$ for all $i$, since the first order temperature field has zero time-average.
We also have Nu$_1$ = 0 for $\mathbf{a} =\mathbf{0}$ since 
Nu$(\mathbf{0}) \equiv$ Nu$_0$ so the remaining terms in (\ref{Nuexpn}) vanish. 
%Thus
%\begin{align}
%   D\mbox{Nu}_i \bigg|_{\mathbf{a} = \mathbf{0}} =  \lim_{\epsilon \to 0} \frac{\mbox{Nu}(\epsilon\mathbf{e}_i)-\mbox{Nu}(\mathbf{0})}{\epsilon} = \mbox{Nu}_1(\epsilon\mathbf{e}_i)-\mbox{Nu}_1(\mathbf{0}) = 0 \;\forall i. \label{DNu1}
%\end{align}
The entries of the Hessian matrix are also computed using the expansion (\ref{Nuexpn}):
\begin{align}
     D^2\mbox{Nu}_{ij} \bigg|_{\mathbf{a} = \mathbf{0}} &= \frac{\partial^2 \mbox{Nu}}{\partial \mathrm{a}_i\partial \mathrm{a}_j}\bigg|_{\mathbf{a} = \mathbf{0}} = \lim_{\epsilon \to 0} \frac{\mbox{Nu}(\epsilon\mathbf{e}_i\!+\!\epsilon\mathbf{e}_j)-\mbox{Nu}(\epsilon\mathbf{e}_i)-\mbox{Nu}(\epsilon\mathbf{e}_j)+\mbox{Nu}(\mathbf{0})}{\epsilon^2} \label{Hessian}\\
    &= \mbox{Nu}_2(\epsilon\mathbf{e}_i\!+\!\epsilon\mathbf{e}_j)-\mbox{Nu}_2(\epsilon\mathbf{e}_i)-\mbox{Nu}_2(\epsilon\mathbf{e}_j). \label{Nu23}
\end{align}
These entries are nonzero in general, so the quadratic term in (\ref{Nuexpn}) is the leading order change to Nu due to the unsteady flow perturbation. Since the Hessian is symmetric it has real eigenvalues, and the optimal small unsteady flow perturbations are those with the largest (i.e. most positive) eigenvalues. Therefore, we compute the Hessian for a range of $\tau$ and Pe, compute the largest eigenvalues in each case, and examine the corresponding eigenvectors (i.e flows). For flows of the form (\ref{Psi}), the Hessian is 2$N$-by-2$N$, with $N = M^2$. We take $M$ = 45. This value is chosen in proportion to $m$ (it is slightly below $m/8$ = 48 when $m = 384$). If $M$ were much larger, the highest degree Chebyshev polynomial used in the modes would not be resolved well on the stretched grids. Each Hessian has 4$M^4 \approx 1.6 \times 10^7$ entries to compute, reduced by a factor of 2 by symmetry, and we compute the Hessian at a large number of $\tau$ and Pe values. We now describe an efficient to method to compute a single Hessian matrix, i.e. the large number of entries given by (\ref{Nu23}) for all $i$ and $j$. 

The sequence of computations for the Hessian is as follows. First, given $\psi_0$, we
solve the zeroth-order problem (\ref{T0}) for $T_0$. Next, we work on obtaining Nu$_2(\epsilon\mathbf{e}_i)$ for $i = 1,\ldots, N$. This means the flow mode coefficient vector
$\mathbf{a} = \epsilon\mathbf{e}_i$ for $i = 1,\ldots, N$, so our flow (\ref{Psi}) has a single nonzero term in the first summand and none in the second, i.e. a cosine function of time. We put this in the form (\ref{psipert}) by setting $f_A$ to column $i$ of $\mathbf{V}$, $f_B$ to $\mathbf{0}$, and $c_1$ to 1. Now we follow the remaining steps in that section to get Nu$_2$: we solve the system (\ref{T1A})--(\ref{T1B}) for ($T_{1A}$, $T_{1B}$) and then solve (\ref{T2s}) for $T_{2s}$ and compute Nu$_2$ from (\ref{Nu2}). To compute Nu$_2(\epsilon\mathbf{e}_i)$ for $i = N+1,\ldots, 2N$, the steps are the same but we now have a single sine term instead of a cosine term in (\ref{Psi}). So we set $f_A$ to $\mathbf{0}$, $f_B$ to column $i$ of $\mathbf{V}$, and $c_1$ to 1. We can get ($T_{1A}$, $T_{1B}$) from the previous computations by the following property of (\ref{T1A})--(\ref{T1B}): if the solution for ($T_{1A}$, $T_{1B}$) is ($g,h$) when ($f_{A}$, $f_{B}$) = ($f,0$), then the solution for ($T_{1A}$, $T_{1B}$) is ($-h,g$) when ($f_{A}$, $f_{B}$) = ($0,f$). We proceed as before, solving for $T_{2s}$ and Nu$_2$ to obtain Nu$_2(\epsilon\mathbf{e}_i)$ for $i = N+1,\ldots, 2N$. Having computed Nu$_2(\epsilon\mathbf{e}_i)$ in (\ref{Nu2}) for all $i$, the same set of values gives Nu$_2(\epsilon\mathbf{e}_j)$ for all $j$.

For Nu$_2(\epsilon\mathbf{e}_i\!+\!\epsilon\mathbf{e}_j)$, we have two unsteady terms in (\ref{Psi}) which may appear in either summand. We put this flow in the form (\ref{psipert}) by setting $c_1 = 2$ and set $f_A$ and $f_B$ according to the four rows of table \ref{Nu2Table}.
\begin{table}[h]
\caption{Choices of $f_A$ and $f_B$ for the computation of Nu$_2(\epsilon\mathbf{e}_i+\!\epsilon\mathbf{e}_j)$ with $i$ and $j$ in different ranges.}\label{Nu2Table}
\centering
\begin{tabular}{c|c|c|c}
$i$ range & $j$ range  & $f_A$ & $f_B$ \\ \hline
$1, \ldots, N$ & $1, \ldots, N$  & $\mathbf{V}_{i} + \mathbf{V}_{j}$ & $\mathbf{0}$ \\ \hline
$1, \ldots, N$ & $N+1, \ldots, 2N$  & $\mathbf{V}_{i}$ & $\mathbf{V}_{j-N}$ \\ \hline
$N+1, \ldots, 2N$ & $1, \ldots, N$  & $\mathbf{V}_{j}$ & $\mathbf{V}_{i-N}$ \\ \hline
$N+1, \ldots, 2N$ & $N+1, \ldots, 2N$ & $\mathbf{0}$ & $\mathbf{V}_{i-N} + \mathbf{V}_{j-N}$ \\ \hline
\end{tabular}
\end{table}
$\mathbf{e}_i$ is a cosine mode in rows 1 and 2 and a sine mode in rows 3 and 4; $\mathbf{e}_j$ is a cosine mode in rows 1 and 3 and a sine mode in rows 2 and 4.
($T_{1A}$, $T_{1B}$) is simply the sum of those for Nu$_2(\epsilon\mathbf{e}_i)$ and Nu$_2(\epsilon\mathbf{e}_j)$, since the right hand side of (\ref{T1A})--(\ref{T1B}) is linear in $f_A$ and $f_B$. We then form the right hand side of (\ref{T2s}), solve it for $T_{2s}$ and compute Nu$_2$ from (\ref{Nu2}).

We can avoid solving the matrix equation (\ref{T2s}) for each of the terms in (\ref{Nu23}) by instead solving an adjoint equation just once, greatly reducing the computing time. To do this, we write equations (\ref{T2s})--(\ref{Nu2}) in discretized form and then combine them:
\begin{align}
    \mathbf{M}_{s} \mathbf{T}_{2s} = \mathbf{b}_{2s} \quad ; \quad \mbox{Nu}_2 = \mathbf{M}_{\mathrm{Nu}} \mathbf{T}_{2s} \quad \rightarrow \quad\mbox{Nu}_2 = \mathbf{r}_{\mathrm{Nu}}^T \mathbf{b}_{2s} \quad ; \quad 
 \mathbf{r}_{\mathrm{Nu}}\equiv \mathbf{M}_{s}^{-T}\mathbf{M}_{\mathrm{Nu}}^T. \label{rNu}
\end{align}
Here $\mathbf{M}_{s}$ is the $m(m-1)$-by-$m(m-1)$ matrix composed of the discretized operators that act on $T_{2s}$ in (\ref{T2s}), now written in discrete form as the vector $\mathbf{T}_{2s}$. The right hand side vector for (\ref{T2s}) is written $\mathbf{b}_{2s}$. $\mathbf{M}_{\mathrm{Nu}}$ is the $1$-by-$m(m-1)$ matrix (or row vector) that corresponds to (\ref{Nu2}) as a mapping from $\mathbf{T}_{2s}$ to Nu. It is a linear mapping formed as a composition of linear mappings: taking the $y$-derivative of $\mathbf{T}_{2s}$ at the two boundaries with finite differences followed by integrating the difference of the two terms using the trapezoidal rule.
In the end we obtain an $m(m-1)$-by-1 column vector $\mathbf{r}_{\mathrm{Nu}}$ that maps the right hand side of (\ref{T2s}) to Nu$_2$ by an inner product.  $\mathbf{r}_{\mathrm{Nu}}$ can be computed easily by solving a single sparse matrix equation, $\mathbf{M}_{s}^{T}\mathbf{r}_{\mathrm{Nu}}=\mathbf{M}_{\mathrm{Nu}}^T$. For the Hessian there are $\sim 10^7$ right hand side vectors for (\ref{T2s}). Instead of solving the matrix equation (\ref{T2s}) $10^7$ times to obtain $\mathbf{T}_{2s}$ and then integrating (\ref{Nu2}) for Nu$_2$, we simply take the inner products of the $\sim 10^7$ right hand sides with the constant vector $\mathbf{r}_{\mathrm{Nu}}$, with is several orders of magnitude faster. The dominant cost then is forming the $\sim 10^7$ right hand side vectors. With this approach the total time to compute a Hessian is about 2 hours in Matlab on a recent 8-core Intel processor, instead of about 6 days if the matrix equation (\ref{T2s}) were solved for each right hand side vector.

We mentioned that only about half the Hessian entries need to be computed because it is a symmetric matrix. In fact, we can reduce the work by another factor of two. As we show in appendix \ref{app:Decomp}, the Hessian is block skew-symmetric:
\begin{align}
    D^2\mbox{Nu} =\left[\begin{array}{cc}
A &B \\ -B&A \\ 
 \end{array}\right] \label{BlockSkewSymm}
\end{align}
and since the Hessian is also symmetric, the blocks $A$ and $B$ are symmetric and skew-symmetric, respectively. Therefore to form the Hessian we only need to compute about half the entries of $A$ and $B$, or about one quarter of the entries of the Hessian.

We have described how the Hessian is computed for flows of the form (\ref{Psi}), with just a single period.
If the flow has unsteady perturbation modes with multiple periods, we show in appendix \ref{app:BlockDiagonal} that the Hessian is block diagonal, with the blocks consisting of the Hessians for each period alone.
Therefore the problem is decoupled, with 
the eigenvalues and eigenvectors given by those for each period alone.  Next, we compute the single-period blocks of the Hessian for various values of the period, and the corresponding eigenvectors and eigenvalues. We focus on cases with positive eigenvalues, so the eigenvectors give flow perturbations that increase Nu.

\section{Results}

\begin{figure}
    \centering
    \includegraphics[width=1\textwidth]{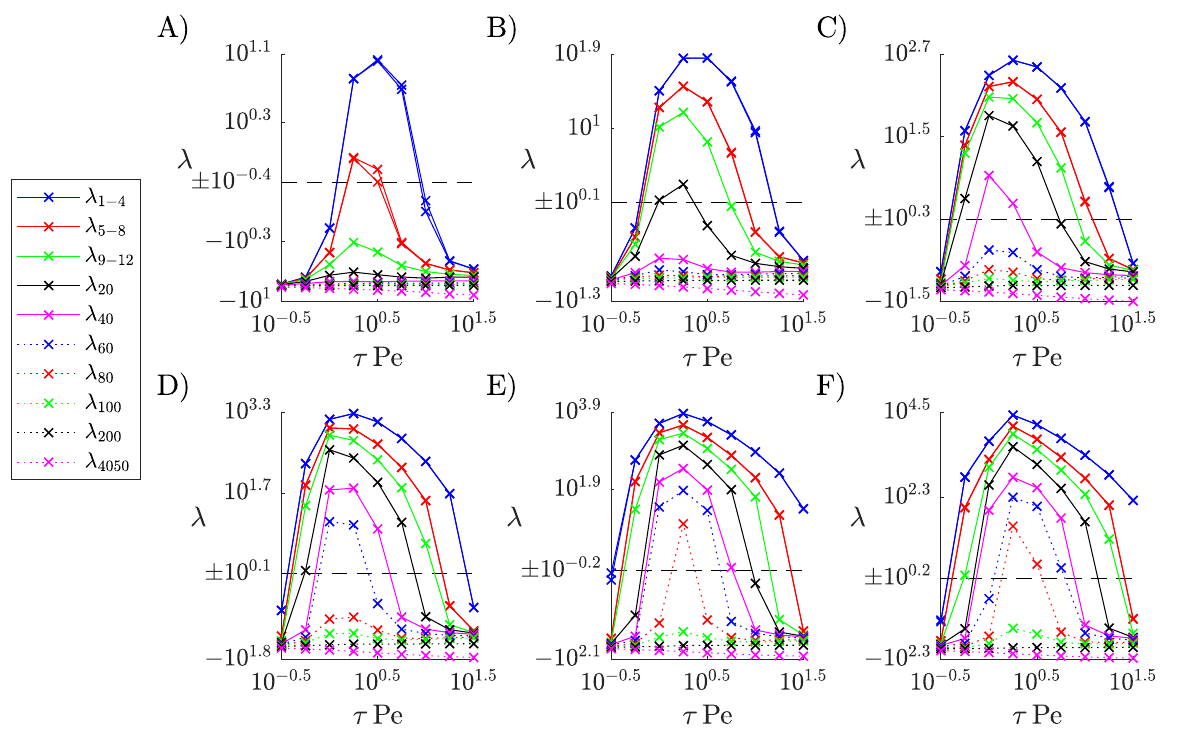}
    \caption{\footnotesize Distributions of Hessian eigenvalues versus $\tau$Pe at six Pe values: A) 2$^9$, B) 2$^{11}$, C) 2$^{13}$, D) 2$^{15}$, E) 2$^{17}$, and F) 2$^{19}$. }
\label{fig:EigenvalueDistributionsFig}
             \vspace{-.10in}
\end{figure}

We compute the Hessian across wide ranges of $\tau$ and Pe values, searching for positive eigenvalues. For each Pe value, we find that positive eigenvalues occur only in a single band of $\tau$ values. In figure \ref{fig:EigenvalueDistributionsFig} we plot the eigenvalues at a fixed Pe in each panel; the Pe values range from $2^9$ to $2^{19}$ as we move from panel A to panel F. We find that the curves align horizontally if we plot them with respect to $\tau$Pe instead of $\tau$. In other words, the band of $\tau$ where positive eigenvalues occur scales approximately as Pe$^{-1}$. Multiple curves are plotted in each panel, and these correspond to particular eigenvalues in the ordering from greatest ($\lambda_1$) to least ($\lambda_{4050}$), labeled by color and line type in the legend. We find that the eigenvalues occur approximately in groups of four (quartets) with almost identical values. Therefore, the solid blue curves used for the top four eigenvalues ($\lambda_1$--$\lambda_4$) almost align in each panel, and likewise for the following quartet, $\lambda_5$--$\lambda_8$, $\lambda_9$--$\lambda_{12}$, etc. We plot a few more eigenvalues (the 20th, 40th, \ldots, 4050th) without listing in the legend the quartets that they belong to. All other eigenvalues lie between the lines that are plotted. 

The quartets can be divided into two pairs, with the eigenvalues agreeing to about 14 digits within each pair, and agreeing to 2--3 digits across the pairs. 
The pairs with 14 digits of agreement are due to the block-skew-symmetric form of the Hessian in equation (\ref{BlockSkewSymm}). If [$v_1$; $v_2$] is an eigenvector of (\ref{BlockSkewSymm}) with eigenvalue $\lambda$, one can show that [-$v_2$; $v_1$] is an eigenvector with the same eigenvalue. Round-off error gives a small difference.

The less exact but still close agreement across the pairs is due to eigenvectors representing flows that are localized near either the top and bottom wall, and relatively small in the center of the channel. One eigenvalue corresponds to a flow near the top wall and the other to almost the same flow near the bottom wall.

A log scale is used for both axes in each panel of figure \ref{fig:EigenvalueDistributionsFig}. Since the eigenvalues are both positive and negative, separate log scales are used on the vertical axis for the positive and negative values and joined together at $\pm 10^c$ for different $c$ in each panel. $c$ is chosen so there are no eigenvalues in the omitted range, $[-10^c, 10^c]$. 

The number of positive eigenvalues at Pe = 2$^9$ (panel A) is at most eight (two quartets), when $\tau$Pe = 10$^{0.25}$--10$^{0.5}$, and increases to between 80 and 100 (20 to 25 quartets) at Pe = 2$^{19}$ (panel F), at the same $\tau$Pe. The peak eigenvalues increase with Pe approximately as power laws, $\lambda_{max}$(Pe) $\sim$ Pe$^{1.33}$ from panels A to C and $\lambda_{max}$(Pe) $\sim$ Pe$^{1}$ from panels 
C to F. Moving from the top quartet of eigenvalues (blue) to the second (red) and third (green), the eigenvalues decrease rapidly in magnitude. The decrease can be fit roughly as a power law, $\sim$(quartet number)$^\alpha$ with $-2 \lesssim \alpha \lesssim -0.8$, depending on Pe and $\tau$.

The smallest eigenvalues are negative and are less of a focus here but they also increase in magnitude with Pe, as $|\lambda_{min}$(Pe)$| \sim$ Pe$^{0.4}$ from panels A to F, and have a slight dependence on $\tau$. The vast majority of eigenvalues are clustered near the minimum, between the $\lambda_{200}$ and $\lambda_{4050}$ lines.

The bandwidth of $\tau$Pe with positive eigenvalues grows with Pe, and at the largest Pe the curves become more asymmetric, rising sharply at small $\tau$Pe and falling off gradually at large $\tau$Pe.

\begin{figure}
    \centering
    \includegraphics[width=1\textwidth]{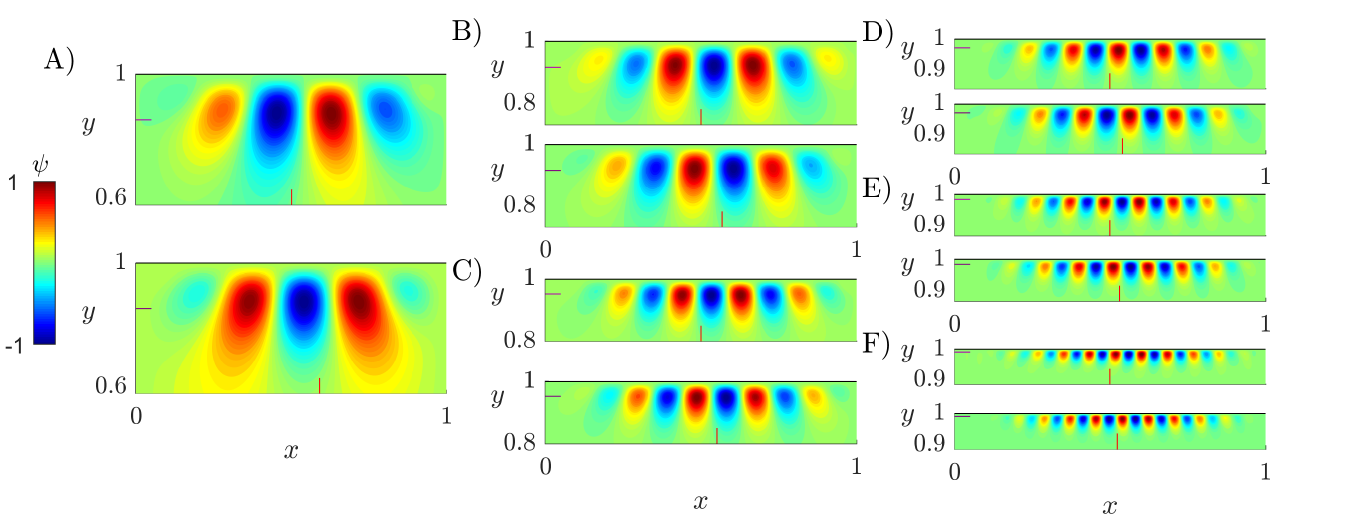}
    \caption{\footnotesize Stream functions corresponding to the Hessian eigenvectors with the largest eigenvalue at $\tau$Pe = 10$^{0.25}$ and six Pe values: A) 2$^9$, B) 2$^{11}$, C) 2$^{13}$, D) 2$^{15}$, E) 2$^{17}$, and F) 2$^{19}$. Within each panel, the top and bottom subplots give the $\cos(2\pi t/\tau)$ and $\sin(2\pi t/\tau)$ components of the stream function.}
    \label{fig:TopEigenfunctionsFig}
             \vspace{-.10in}
\end{figure}

The main interest of this study is the unsteady flows that are most beneficial for heat transfer. For small unsteady perturbations, these are the flows corresponding to the largest positive eigenvalues. For the largest eigenvalues at $\tau$Pe = 10$^{0.25}$ in each of the panels A--F of figure \ref{fig:EigenvalueDistributionsFig}, we present the flows of the corresponding eigenvectors in panels A--F of figure \ref{fig:TopEigenfunctionsFig}. This value of $\tau$Pe maximizes the blue curves in figure \ref{fig:EigenvalueDistributionsFig}C--F, and is slightly below the maxima in panels A and B.

Each panel shows the stream function of the flow in two subpanels. The top and bottoms subpanels show the cosine and sine components, $f_A$ and $f_B$ in equation (\ref{psipert}). The overall magnitudes are arbitrary since the flows correspond to eigenvectors. The stream functions are nonzero only in a small region near the top wall, shown in the panels. As discussed previously there is another eigenvector with nearly the same eigenvalue that gives nearly the same flow at the bottom wall.  The flows consist of sequences of alternating eddies whose sizes and distances from the wall decrease as Pe increases from A--F. The flows in the top and bottom subpanels, $f_A$ and $f_B$ in equation (\ref{psipert}), are almost the same but shifted by one quarter of a horizontal period. We can approximate them as
\begin{align}
f_A(x,y) \approx F(x)G(y)\cos\left(\frac{2\pi x}{\lambda}\right); \quad f_B(x,y) \approx F(x)G(y)\sin\left(\frac{2\pi x}{\lambda}\right) 
\end{align}
with $F(x)$ an envelope function that decays smoothly from 1 in the middle ($x = 1/2$) to 0 at the left and right boundaries ($x = 0, 1$), and $G(y)$ an envelope function that decays from 1 at the top wall to zero outside of a boundary layer thickness that depends on Pe. Then the total perturbation flow is 
\begin{align} 
\psi_1 = f_A(x,y)\cos\left(\frac{2\pi t}{\tau}\right)+f_B(x,y)\sin\left(\frac{2\pi t}{\tau}\right) \approx
F(x)G(y)\cos\left(2\pi\left(\frac{x}{\lambda}-\frac{t}{\tau}\right)\right)
\end{align}
a traveling wave multiplied by envelopes. We now compare the speed of the traveling wave to the maximum horizontal speed of the steady optimal flow, $u_0(y)$, shown in figure \ref{fig:ChannelSteadyOptFig}. The flow speed rises to nearly its maximum over the boundary layer, whose edge can be estimated by the $y$ location where the steady velocity is 99 percent of its maximum. This is marked by small horizontal lines at the left edge of each subpanel. The centers of the eddies in panel A are close to the boundary layer edge, while the eddies in panel F are mostly outside the boundary layer. Therefore in all panels the eddies occur where the steady flow speed is about its maximum.  To compare the traveling wave speed to the maximum steady flow speed, we place small vertical lines at $x$ = 1/2 at the bottom of the upper subpanels. In the lower subpanels, we displace the vertical lines horizontally by the amount they would move under the maximum steady flow as $t$ changes from 0 (when the flow is given by the cosine component) to $\tau$/4 (when the flow is given by the sine component). The horizontal distance is $\tau$/4 times the maximum steady flow speed. Comparing the vertical lines in the upper and lower subpanels in A--F, we see that they have about the same position relative to the eddies. Therefore the eddy velocity (the traveling wave velocity) is about the same as the steady flow speed. 

\begin{figure}
    \centering
    \includegraphics[width=0.9\textwidth]{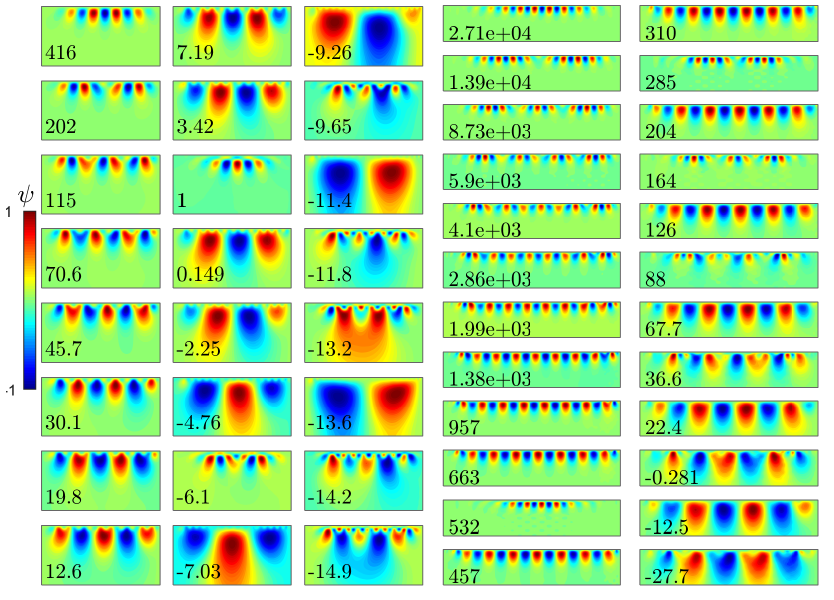}
    \caption{\footnotesize Stream functions corresponding to the Hessian eigenvectors from the top 24 quartets of eigenvalues at $\tau$Pe = 10$^{0.25}$ and two Pe values: 2$^{13}$ (three leftmost columns) and 2$^{19}$ (two rightmost columns). Only the $\cos(2\pi t/\tau)$ components of the stream functions are shown. For each Pe (each group of three/two columns), the stream functions are ordered by eigenvalue (listed at the lower left corner of each panel), starting at the top of the leftmost column and proceeding downward and then similarly through the column(s) to the right. Only $0.6 \leq y \leq 1$ is shown for the three leftmost columns, and only $0.8 \leq y \leq 1$ is shown for the two rightmost columns.}
    \label{fig:MultipleEigenfunctionsFig}
             \vspace{-.10in}
\end{figure}

Next, we compare the flows that correspond to different eigenvalues. In figure \ref{fig:MultipleEigenfunctionsFig} we show the flows for eigenvalues in the top 24 quartets for two different Pe, 2$^{13}$ and 2$^{19}$, again at $\tau$Pe = 10$^{0.25}$. These are the eigenvalues that range from the solid blue line to the dotted green line in panels C and F in figure \ref{fig:EigenvalueDistributionsFig}. The flows are arranged in order of decreasing eigenvalue (values listed on each subpanel) moving downward within each column, starting from the left column and moving rightward. The flows for Pe = 2$^{13}$ are in the three leftmost columns, and those for Pe = 2$^{19}$ are in the rightmost two. The top flows at each Pe were shown in panels C and F of the previous figure, and correspond to a single chain of alternating eddies. The second-, third-, and fourth-best flows are similar but are broken into two, three, and four clusters of eddies respectively. The four clusters are easier to recognize at Pe = 2$^{19}$ where there are more eddies. After this first group of small clusters of eddies, other patterns appear which are larger and therefore easier to see in the left columns, Pe = 2$^{13}$. From the fourth flow onward, the eddies often adopt a scalloped shape near the boundary. Here the indented streamlines encapsulate smaller closed eddies near the wall. These smaller eddies are sometimes strong enough to be colored red and blue, as for $\lambda$ = -14.9 and -9.65. In such cases the stream function magnitudes are comparable in the largest and smallest eddies. Since the eddy flow velocity scales as the stream function magnitude divided by the eddy size, the flow velocity is much larger in the small eddies in these cases. These patterns recall the branching flow structures near walls in optimal steady flows between hot and cold walls \cite{motoki2018maximal,souza2020wall,kumar2022three,alben2023transition,alben2024optimal}.

As secondary layers of eddies develop near the wall, the largest eddies generally become larger and extend further from the wall.
At both Pe, there is no obvious change in the types of flows when the eigenvalues pass through zero and become negative. 

We have shown bands of flows with large positive eigenvalues, which increase the Nusselt number as the square of the unsteady flow amplitude, in the limit of small amplitude. We now investigate the effectiveness of these flows when the unsteady component has moderate and large amplitude.  Flows of the form (\ref{Psi}) have power Pe for arbitrary real $\mathbf{a}$. As $\|\mathbf{a}\|$ grows, the unsteady part grows at the expense of the steady part. As $\|\mathbf{a}\| \to \infty$, we have only the unsteady part, which obeys no-slip boundary conditions. With no inflow or outflow, the temperature boundary conditions given below equation (\ref{AdvDiff}) remain mathematically valid but describe a different physical problem: 
flow in a closed box with an insulated wall at the right boundary and fixed temperature on the other three sides. Then heat leaves the domain by conduction through the left (cold) wall and nearby regions of the top and bottom walls. We find in our simulations, described next, that the flows become ineffective well before this limit.  

We simulate the unsteady advection-diffusion equation (\ref{AdvDiff}) with flows of the form (\ref{Psi}) where $\mathbf{a}$ is an eigenvector with either the largest or the fifth largest eigenvalue, i.e. from the first or second quartets. These are the eigenvalues on the blue and red curves in figure \ref{fig:EigenvalueDistributionsFig}. For the six Pe and nine $\tau$Pe in that figure, we set $\mathbf{a}$ to $\mathbf{v}_1$ and $\mathbf{v}_5$, the eigenvectors corresponding to $\lambda_1$ and $\lambda_5$, in all the cases where these eigenvalues are positive. In such cases, we vary $\epsilon \equiv$ $\|\mathbf{a}\|$ from 10$^{-1.5}$ to 10$^{0.75}$. At the lower end of this range, the unsteady flow is small relative to the steady part, while at the upper end the unsteady flow may be larger at certain times near the upper and lower walls. However, the steady flow is always dominant near the inflow and outflow boundaries, so the inflow/outflow boundary conditions are valid.

\begin{figure}
    \centering
    \includegraphics[width=0.8\textwidth]{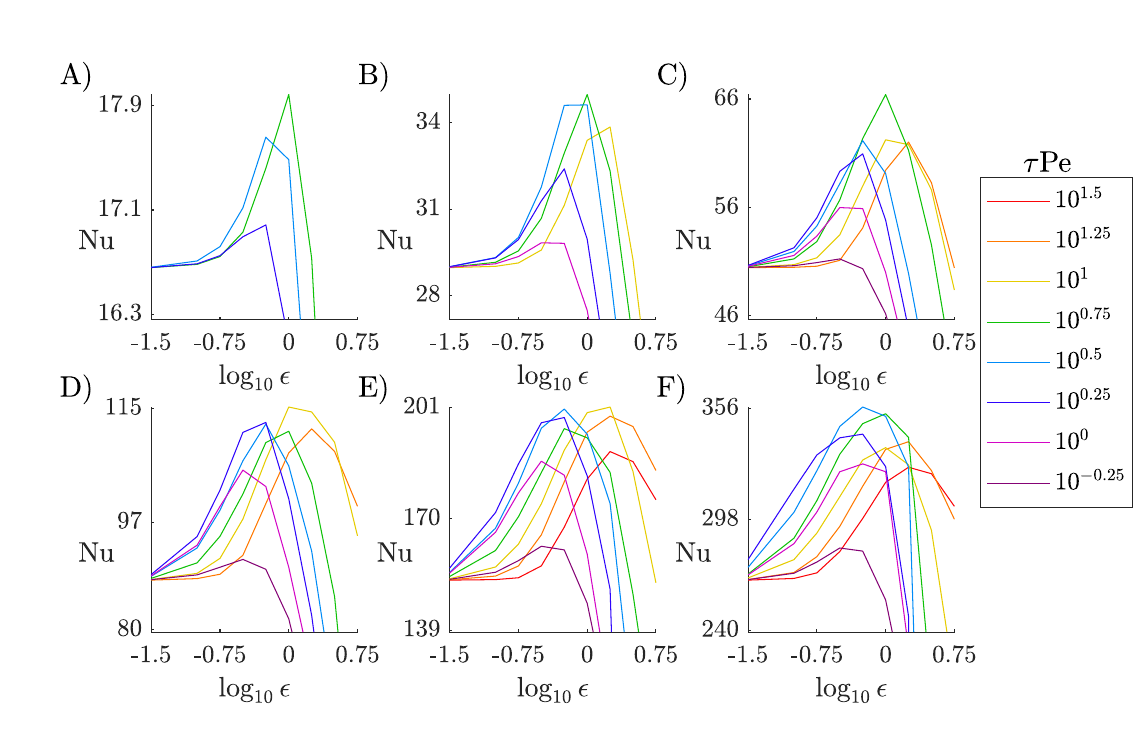}
    \caption{\footnotesize Nu versus perturbation amplitude $\epsilon$, for flows corresponding to the largest Hessian eigenvalues at six Pe values: A) 2$^9$, B) 2$^{11}$, C) 2$^{13}$, D) 2$^{15}$, E) 2$^{17}$, and F) 2$^{19}$. At each Pe, curves are plotted for a subset of the $\tau$Pe values listed at right, those for which the largest eigenvalue is positive.}
    \label{fig:CNMeanNuFigure1}
             \vspace{-.10in}
\end{figure}

We simulate equation (\ref{AdvDiff}) as an initial value problem, starting from the initial condition given by (\ref{Texpn}) up to $O(\epsilon^2)$ and evaluated at $t = 0$. We use the Crank-Nicolson method for time stepping, and the spatial derivatives are discretized as previously, with second-order finite differences on the stretched grids. We simulate the flows up to $t = 20\tau$, by which time the flows have reached a periodic steady state for several periods in all the cases that increase Nu over the steady value. Some cases with relatively large $\epsilon$ do not reach the periodic state, but they also do not increase Nu and are not considered further.

For the largest eigenvalue $\lambda_1$, the steady-state time-averaged Nu for the simulations are plotted in figure \ref{fig:CNMeanNuFigure1}. Each panel corresponds to the same panel in figure \ref{fig:EigenvalueDistributionsFig}, with the eigenvalue following the blue curves. In figure \ref{fig:CNMeanNuFigure1}A only three curves are shown, because the blue curve in figure \ref{fig:EigenvalueDistributionsFig}A exceeds zero at only three $\tau$Pe values. More curves are shown in the other panels, corresponding $\lambda_1 > 0$ at more $\tau$Pe. By plotting (Nu$-$Nu$_0$) versus $\epsilon$ on log-log axes (not shown), one finds that all the curves rise quadratically with $\epsilon$ for small $\epsilon$, with a curvature given by the corresponding $\lambda_1$, as expected. In figure \ref{fig:CNMeanNuFigure1} log-linear axes are used instead, so Nu$_0$ (the values that the curves approach at the left ends of the plots) and the peak Nu can both be seen.

In each panel of figure \ref{fig:CNMeanNuFigure1}, the peak Nu occur at moderately large $\epsilon$, 10$^{-0.25}$--10$^{0.25}$. They occur on the light blue, green, or yellow curves, i.e. with $\tau$Pe = 10$^{0.5}$ -- 10$^{1}$.
These are generally longer periods than the ones that give the largest eigenvalues in figure 
\ref{fig:EigenvalueDistributionsFig}, which were $\tau$Pe = 10$^{0.25}$ and 10$^{0.5}$.
One can see that the curves that rise most rapidly at small $\epsilon$ in figure \ref{fig:CNMeanNuFigure1} (light blue in A--B and dark purple in C--F) do not give the peak Nu. The largest increases in Nu relative to Nu$_0$ range from 7\% in panel A to 30-35\% in panels C--F. In most cases, curves with larger $\tau$Pe peak at larger $\epsilon$, so lower perturbation frequencies work well with larger perturbation amplitude and higher frequencies with smaller amplitudes. Large increases in Nu can be obtained with a wide range of frequencies, particularly at large Pe, when the optimal amplitude is chosen.

\begin{figure}
    \centering
    \includegraphics[width=0.8\textwidth]{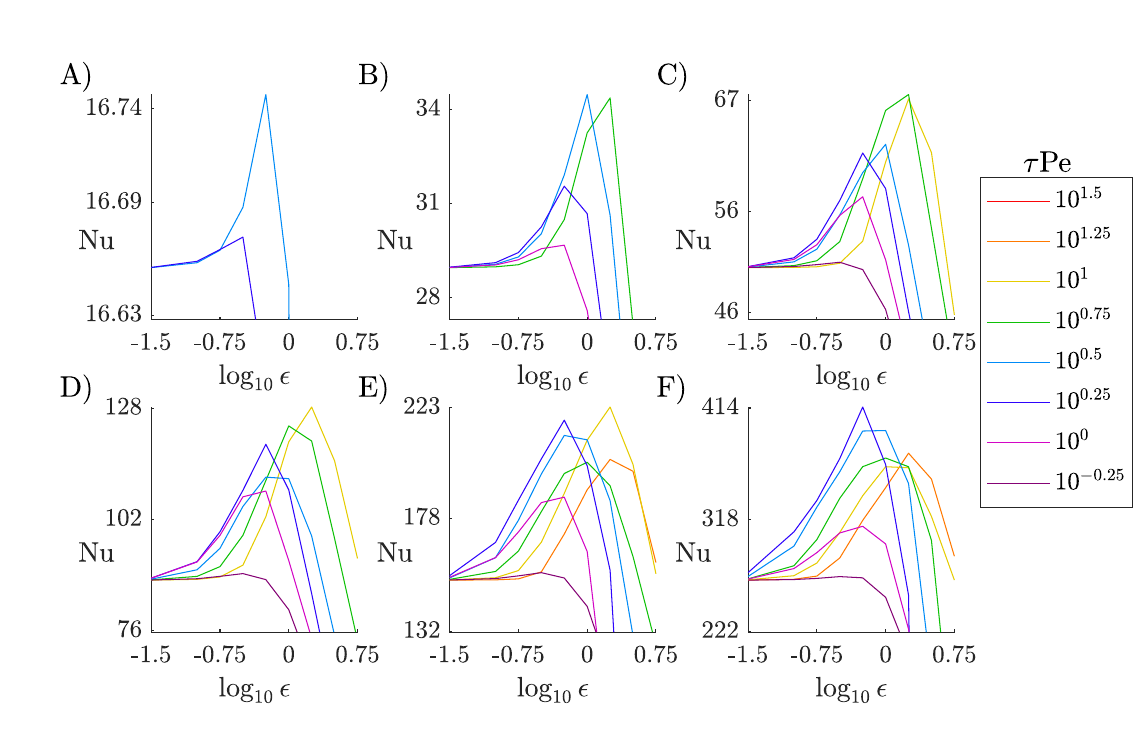}
    \caption{\footnotesize Same quantities as in figure \ref{fig:CNMeanNuFigure1} but for the fifth largest eigenvalue (the second quartet).}
    \label{fig:CNMeanNuFigure2}
             \vspace{-.10in}
\end{figure}

Figure \ref{fig:CNMeanNuFigure2} plots Nu versus perturbation amplitude for $\lambda_5$, i.e. the second quartet, the red curves in figure \ref{fig:EigenvalueDistributionsFig}. The panels have slightly fewer curves than previously, because some $\tau$Pe had only one positive eigenvalue, not two. The peak Nu again occur at $\epsilon$ = 10$^{-0.25}$--10$^{0.25}$. They occur at a similar but somewhat wider range of $\tau$Pe, 10$^{0.25}$ -- 10$^{1}$. 
The dependence of the peaks on $\tau$Pe is complicated at large Pe, panels D--F. At first the peaks rise with $\tau$Pe, reaching a maximum at 10$^{0.25}$, the dark purple curves. They then decrease and increase again, reaching a second maximum at 10$^{1}$ or 10$^{1.25}$ (in F), the yellow or orange curves.
The same phenomenon also occurred in the previous figure but was not discussed.

The heat transfer enhancement Nu$-$Nu$_0$ is very small in figure \ref{fig:CNMeanNuFigure2}A, where the eigenvalues happen to be very small (but positive). By contrast, panels B and C show peak Nu$-$Nu$_0$ values similar to figure 
\ref{fig:CNMeanNuFigure1}B and C, while panels D--F show significantly larger 
peak Nu$-$Nu$_0$ than figure 
\ref{fig:CNMeanNuFigure1}D--F. Thus the second quartet gives larger heat transfer enhancement than the first at large Pe, up to 44--56\% in panels D--F.

\begin{figure}
    \centering
    \includegraphics[width=1\textwidth]{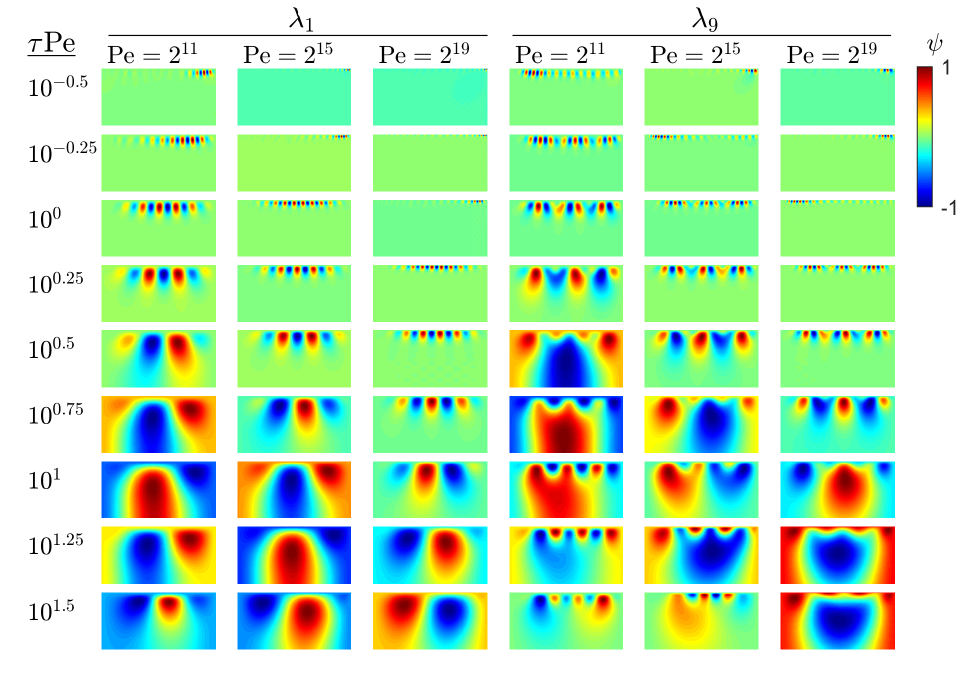}
    \caption{\footnotesize Stream functions corresponding to the Hessian eigenvectors from the best and third-best quartets of eigenvalues, labeled $\lambda_1$ and $\lambda_9$ respectively. Each row corresponds to one of nine $\tau$Pe values (labeled at the left) and each column corresponds to one of three Pe values (labeled at the top).
    Only the $\cos(2\pi t/\tau)$ components of the stream functions are shown, and only in the range $0.5 \leq y \leq 1$.}
    \label{fig:EigenfunctionsVsTauPeFig}
             \vspace{-.10in}
\end{figure}

In figure \ref{fig:EigenvalueDistributionsFig} we found that the eigenvalues have a peak near $\tau$Pe = $10^{0.25}$ but are positive and large over a broad range, broader at larger Pe. In figures \ref{fig:TopEigenfunctionsFig} and \ref{fig:MultipleEigenfunctionsFig} we showed the flows for the top eigenvalues at 
$\tau$Pe = $10^{0.25}$ only. However, in figures \ref{fig:CNMeanNuFigure1} and \ref{fig:CNMeanNuFigure2} we showed Nu across a wide range of $\tau$Pe and found that the maximum Nu occur at $\tau$Pe in the range 10$^{0.5}$--10$^{1}$, larger than $\tau$Pe = $10^{0.25}$ where the small-amplitude peak of Nu, corresponding to the eigenvalues, usually occurred.
To study the flows at the $\tau$Pe where the maximum Nu occur in figures \ref{fig:CNMeanNuFigure1} and \ref{fig:CNMeanNuFigure2}, we show optimal unsteady perturbation flows {\it across} $\tau$Pe in figure \ref{fig:EigenfunctionsVsTauPeFig}. The first three columns are flows for the top eigenvalue $\lambda_1$, including all the $\tau$Pe in figure \ref{fig:CNMeanNuFigure1} and half the Pe, those for panels B, D, and F. We see a clear transition in the flow structure as $\tau$Pe increases from top to bottom. At the top the flows consist of very small eddies concentrated very close to the boundary, and increasingly near the outflow boundary. As $\tau$Pe increases (and as Pe decreases), the eddies become larger and the chains are more centered horizontally. For $\tau$Pe $\gtrsim$ 10$^{0.75}$ the eddies are less localized near the wall and typically fill the whole flow domain. 

In the right three columns, the same quantities are shown for $\lambda_9$, an eigenvalue from the third quartet, for comparison. At 
$\tau$Pe = 10$^{0}$ and neighboring values one can see that the eddies occur in three clusters instead of one, the pattern discussed previously with the sequence of eigenmodes shown in figure \ref{fig:MultipleEigenfunctionsFig}.
Otherwise, the flow patterns and transition with $\tau$Pe are generally the same as for the first quartet, except some modes for $\lambda_9$ are clustered near the inflow side of the top wall instead of the outflow side.

In all of the columns, some of the flows correspond to negative eigenvalues, e.g. the top row, $\tau$Pe = 10$^{-0.5}$, but most have positive eigenvalues. The cases with negative eigenvalues tend to occur at the smallest or largest $\tau$Pe and the smallest Pe, and can be specified by where the solid blue and green lines in figure \ref{fig:EigenvalueDistributionsFig}B, D, and F lie below the dashed lines.

The main takeaway message of figure 
\ref{fig:EigenfunctionsVsTauPeFig} is that the flow perturbations change from localized at the boundary to channel-filling at 
$\tau$Pe $\approx$ 10$^{0.5}$--10$^{1}$, which is the range that gives the peak Nu values at large amplitude in figures \ref{fig:CNMeanNuFigure1} and \ref{fig:CNMeanNuFigure2}. Therefore, both types of flows---wall-bounded with high frequency and channel-filling with low frequency---can be comparably effective at increasing Nu.
We now show examples of the temperature fields that correspond to both types of flows.

\begin{figure}
    \centering
    \includegraphics[width=0.8\textwidth]{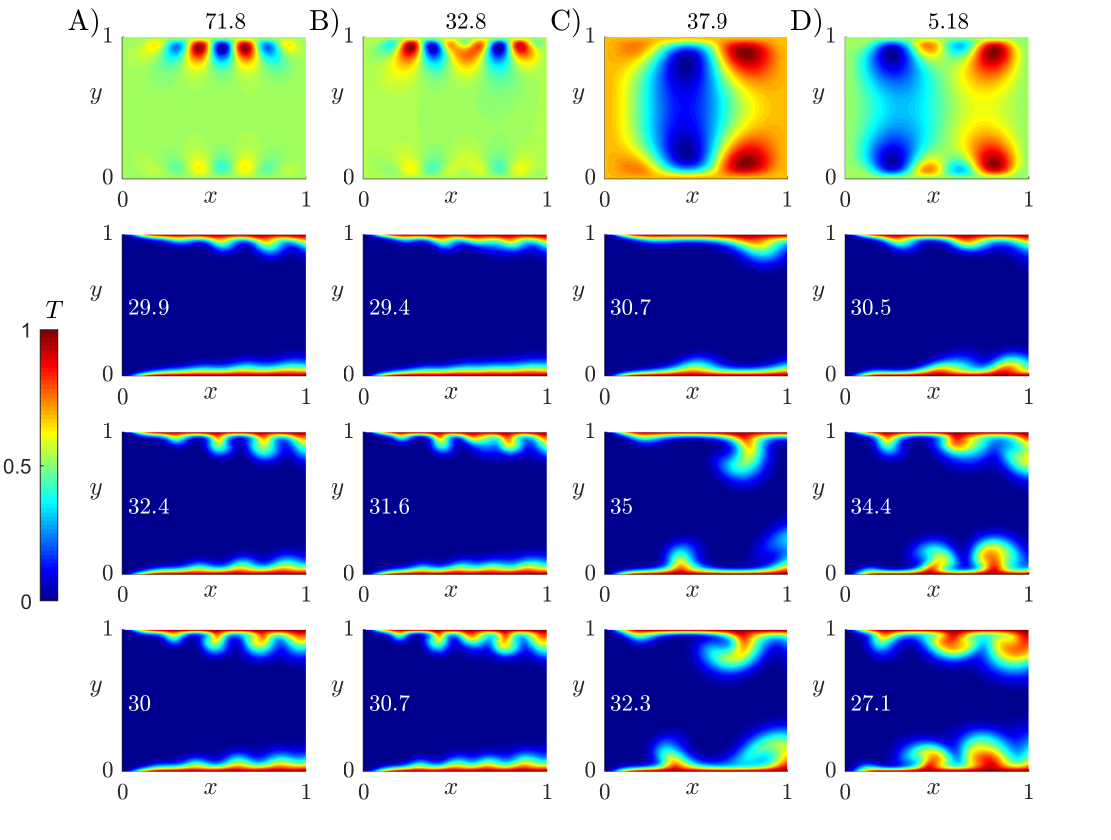}
    \caption{\footnotesize At Pe = 2$^{11}$, flows and temperature fields for the largest and fifth largest eigenvalues. Columns (A) and (B) correspond to a medium period, $\tau$Pe = 10$^{0.25}$ while (C) and (D) correspond to a larger period, $\tau$Pe = 10$^{0.75}$. The first columns of each pair (A and C) correspond to the largest eigenvalue, and the second columns (B and D) correspond to the fifth largest eigenvalue. The top panel of each column is the flow for the corresponding eigenmode, with the eigenvalue listed above. The next three panels, moving downward, plot the temperature fields from the solution of the unsteady advection-diffusion equation with the given flow perturbation at three increasing amplitudes ($\epsilon$). The $\epsilon$ values are: A,B) $10^{-0.75}, 10^{-0.25}, 10^{0}$; C) $10^{-0.5}, 10^{0}, 10^{0.25}$; D) $10^{-0.25}, 10^{0.25}, 10^{0.5}$. The temperature fields are plotted at time $t = 20 \tau$, after a periodic steady state has been reached. The time-averaged Nu is labeled at the left side of each temperature field.}
    \label{fig:CNTemperature11Fig}
             \vspace{-.10in}
\end{figure}

In figure \ref{fig:CNTemperature11Fig} we show flows and temperature fields 
at two values of $\tau$Pe that straddle the transition from local to global flows. The flows are shown in the top panels of each column, at a medium $\tau$Pe, corresponding to a local flow, in columns A and B, and a large $\tau$Pe, corresponding to a global flow,
in columns C and D. The first of each pair of columns (A and C) give the flow and temperature fields for $\lambda_1$, the top quartet. The second pair of columns (B and D) give the same quantities for $\lambda_5$, the second quartet. Below the top panels we show three rows of temperature field snapshots, for three values of the perturbation amplitude $\epsilon$: small, medium, and large (relative to the steady flow), as one moves downward. The middle of the three $\epsilon$ gives the peak Nu value. For all of these cases, Pe = 2$^{11}$, corresponding to the dark purple and green curves in figures \ref{fig:CNMeanNuFigure1}B and \ref{fig:CNMeanNuFigure2}B, at three $\epsilon$ that include the peaks. 

For all the cases in figure \ref{fig:CNTemperature11Fig}, Nu$_0$ = 29. Flows A and B give peak increases of 12 and 9\% respectively, versus 21 and 19\% for flows C and D, which have much smaller eigenvalues than A and B (listed at the top of the columns). The smaller eddies of A and B correspond to smaller and more numerous temperature plumes emanating from the top wall (and similar but weaker eddies and plumes at the bottom wall). By contrast, flows C and D have up-down symmetric streamlines and the large eddies create significant plumes at both walls. However, there is also an energetic cost to having a nonuniform flow across the entire channel. For all of these flow perturbations, the basic mechanism for increasing Nu involves the eddies bringing cold fluid toward the wall while moving hot fluid away from it at certain locations. The local temperature boundary layer thickness decreases and increases at these locations, respectively, but the net effect is to decrease the boundary layer thickness (and increase Nu) except in the lower right corner. Here $\epsilon = 10^{0.5}$, and the unsteady flow is much stronger relative to the steady flow than in the panel above. The plumes of hot fluid are also much wider, resulting in a net increase in the boundary layer thickness. In general, the optimal temperature fields (those in the middle row) are qualitatively similar to those above and below it, but represent an optimal balance of mixing by the unsteady part and advection by the steady part. 

\begin{figure}
    \centering
    \includegraphics[width=0.8\textwidth]{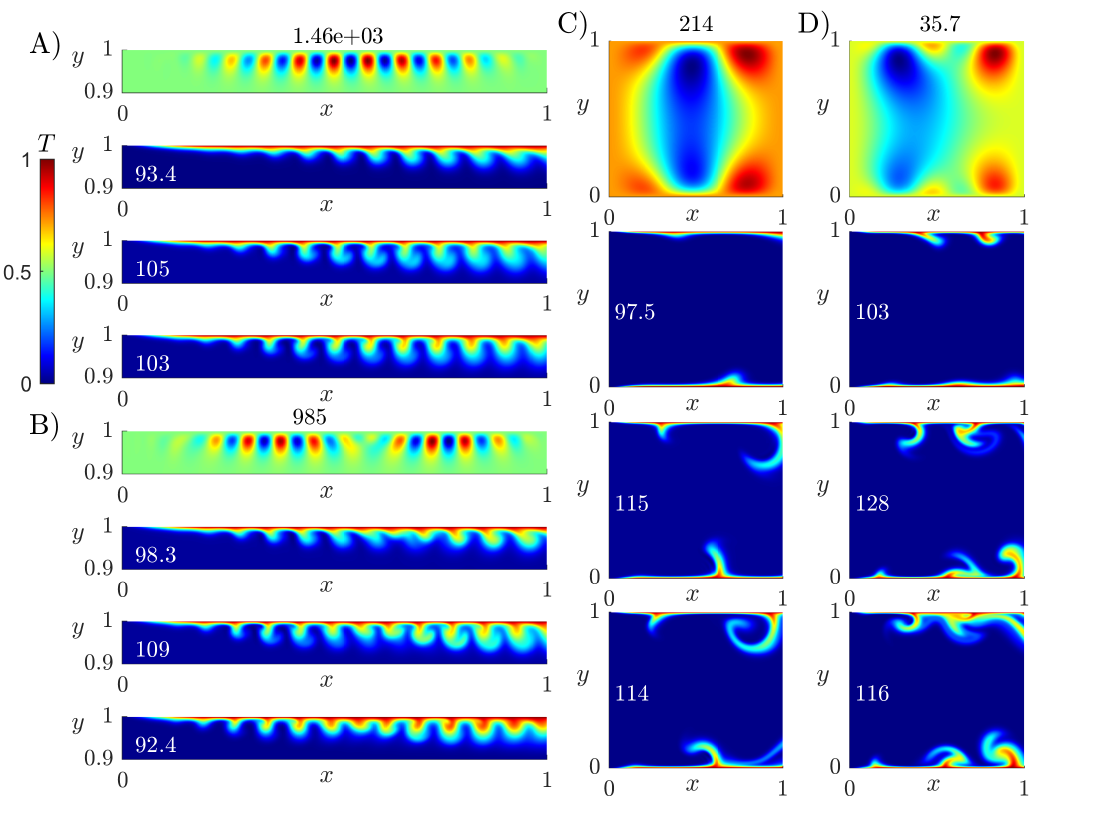}
    \caption{\footnotesize The same quantities as in figure \ref{fig:CNTemperature11Fig}, but with
    Pe increased to 2$^{15}$ and $\tau$Pe = $10^{0}$ in columns A and B and $10^1$ in columns C and D. The $\epsilon$ values are also different: A) $10^{-1}, 10^{-0.5}, 10^{-0.25}$; B) $10^{-0.75}, 10^{-0.25}, 10^{0}$; C) $10^{-0.5}, 10^{0}, 10^{0.25}$; D) $10^{-0.25}, 10^{0.25}, 10^{0.5}$.}
    \label{fig:CNTemperature15Fig}
             \vspace{-.10in}
\end{figure}

Figure \ref{fig:CNTemperature15Fig} shows the same quantities for Pe increased to 2$^{15}$, and $\tau$Pe set to $10^{0}$ in columns A and B (stacked vertically) and $10^1$ in columns C and D. The cases correspond to the magenta and yellow curves in figures \ref{fig:CNMeanNuFigure1}D and \ref{fig:CNMeanNuFigure2}D, at three $\epsilon$ that include the peaks. In this figure, Nu$_0$ = 88, so flows A and B give peak Nu increases of 19 and 24\% respectively, versus 31 and 45\% for flows C and D. Again, the eigenvalues (at the top of each column) are much larger for A--B than for C--D. In columns A and B, there are many more eddies and hot plumes than in columns of A and B of the previous figure. In the optimal cases (the middle temperature fields), the plumes extend farthest from the wall but the outer portions of the plumes are cooler than in the cases below. Hence the very hot (red) part of the boundary layer is thinner in the middle temperature fields, resulting in larger Nu. Columns C and D show the results of the large, low-frequency eddies, slightly less symmetric than in columns C and D the previous figure but having almost the same form. The temperature fields in figure \ref{fig:CNTemperature15Fig}C and D have plumes in nearly the same places as figure \ref{fig:CNTemperature11Fig}C and D, but they and the wall boundary layers are thinner due to the larger Pe. Nu is almost the same for the last two panels of column C even though $\tau$ differs by a factor of 10$^{0.25} \approx 1.8$. There is a larger difference between Nu for last two panels in column D, which may be due to the larger extent of the hot plume along the top wall in the bottom panel. 

\section{Conclusions}

We have computed optimal {\it unsteady} perturbations to optimal steady unidirectional flows in a channel. In the limit of small perturbations, the leading-order change in the heat transfer (Nu) is quadratic in the perturbation size. We expanded the perturbation in a set of flow modes that are orthogonal with respect to the power integral (Pe$^2$). Increases in heat transfer then correspond to positive eigenvalues of the Hessian matrix, the second derivatives of Nu with respect to the mode coefficients. Perturbations with different periods ($\tau$) are decoupled and can be considered separately (i.e. the Hessian is block diagonal). We gave an efficient method to compute the $\sim 10^7$ entries of the Hessian at a range of $\tau$ and Pe, and then we searched for Hessians with positive eigenvalues.

We found that positive eigenvalues occur in certain bands of $\tau$ that scale as Pe$^{-1}$. The largest eigenvalues (actually quartets of eigenvalues) correspond to chains of eddies near the walls that move as traveling waves, at a speed close to that of the steady background flow. The next largest eigenvalues give eddy chains subdivided into clusters of two, three, four, etc. Subsequent eigenvalues give eddies with multiple scales, with the smallest adjacent to the wall. The largest eddies are eventually the size of the entire flow domain. The eddy size decreases with increasing Pe and with decreasing $\tau$. At a crossover $\tau$ (i.e. $\tau$Pe = $O(1)$), the eddies corresponding to the largest eigenvalues change from localized near the wall to domain-filling, with smaller eddies near the wall.

Although the unsteady flows are derived in the small amplitude limit, they give large increases in heat transfer for a given power at large amplitude also---up to about 56\% at the largest Pe studied here, 2$^{19}$. Larger increases occur with the second quartet of eigenmodes than the first, despite having much smaller eigenvalues, and still larger increases may occur with subsequent eigenmodes, not studied in detail here. We obtain large increases in Nu across a range of flow periods $\tau$, and the largest increases occur at larger $\tau$ than those that give the largest eigenvalues. These large $\tau$ correspond to eddies that range up to the domain scale. Hence we obtain heat transfer enhancement by a variety of unsteady flow perturbations, including a wide range of length and time scales.

\section*{Acknowledgements}
\vspace{-.25cm}
\noindent S.A. acknowledges support from the NSF-DMS Applied Mathematics program,
award number DMS-2204900.

\appendix
\section{The Hessian is block skew-symmetric \label{app:Decomp}}
In order to show the Hessian is block skew-symmetric as in equation (\ref{BlockSkewSymm}), we need to show
\begin{align}
\mbox{Nu}_2&(\epsilon\mathbf{e}_i\!+\!\epsilon\mathbf{e}_j)-\mbox{Nu}_2(\epsilon\mathbf{e}_i)-\mbox{Nu}_2(\epsilon\mathbf{e}_j) = D^2\mbox{Nu}_{i,j} \nonumber\\
&= D^2\mbox{Nu}_{i+N,j+N} =\mbox{Nu}_2(\epsilon\mathbf{e}_{i+N}\!+\!\epsilon\mathbf{e}_{j+N})-\mbox{Nu}_2(\epsilon\mathbf{e}_{i+N})-\mbox{Nu}_2(\epsilon\mathbf{e}_{j+N}), \quad i,j = 1, \ldots, N \label{Ablock}
\end{align}
which implies the two blocks labeled $A$ in (\ref{BlockSkewSymm}) are the same and 
\begin{align}
\mbox{Nu}_2&(\epsilon\mathbf{e}_{i+N}\!+\!\epsilon\mathbf{e}_j)-\mbox{Nu}_2(\epsilon\mathbf{e}_{i+N})-\mbox{Nu}_2(\epsilon\mathbf{e}_j) = D^2\mbox{Nu}_{i+N,j} \nonumber\\
&= -D^2\mbox{Nu}_{i,j+N} =-\mbox{Nu}_2(\epsilon\mathbf{e}_{i}\!+\!\epsilon\mathbf{e}_{j+N})+\mbox{Nu}_2(\epsilon\mathbf{e}_{i})+\mbox{Nu}_2(\epsilon\mathbf{e}_{j+N}), \quad i,j = 1, \ldots, N \label{Bblock}
\end{align}
which implies $B$ is the same in the blocks labeled $B$ and $-B$ in (\ref{BlockSkewSymm}).

First we show (\ref{Ablock}).  For $\mbox{Nu}_2(\epsilon\mathbf{e}_i)$, $i = 1, \ldots, N$ we have a perturbation that is a cosine function of time, with ($f_{A}$, $f_{B}$) = ($\mathbf{V}_i,\mathbf{0}$) (in discretized form). Let the corresponding solution for ($T_{1A}$, $T_{1B}$) be ($\mathbf{g}_i,\mathbf{h}_i$) (also in discretized form). Then the right hand side of equation (\ref{T2s}) for $T_{2s}$ is 
\begin{align}
    \mathbf{b}_{2s}(\epsilon\mathbf{e}_i) = - \frac{1}{2}\mathbf{D}_y\mathbf{V}_i\cdot\mathbf{D}_x\mathbf{g}_i + \frac{1}{2}\mathbf{D}_x\mathbf{V}_i\cdot\mathbf{D}_y\mathbf{g}_i+ \frac{1}{2}\mathbf{D}_y\mathbf{\Psi}_0 \cdot\mathbf{D}_x\mathbf{T}_0 - \frac{1}{2}\mathbf{D}_x\mathbf{\Psi}_0\cdot\mathbf{D}_y\mathbf{T}_0
\end{align}
where $\mathbf{D}_x$ and $\mathbf{D}_y$ are discrete first $x$ and $y$ derivative operators (finite difference matrices), $\mathbf{T}_0$ is the discrete (vector) form of $T_0$, and $\cdot$ is the componentwise product of vectors.

For $\mbox{Nu}_2(\epsilon\mathbf{e}_j)$, $j = 1, \ldots, N$, mode $j$ is also a perturbation with a cosine function of time, so we have ($f_{A}$, $f_{B}$) = ($\mathbf{V}_j,\mathbf{0}$). The calculation proceeds as in the previous paragraph but with $j$ in place of $i$:
\begin{align}
    \mathbf{b}_{2s}(\epsilon\mathbf{e}_j) = - \frac{1}{2}\mathbf{D}_y\mathbf{V}_j\cdot\mathbf{D}_x\mathbf{g}_j + \frac{1}{2}\mathbf{D}_x\mathbf{V}_j\cdot\mathbf{D}_y\mathbf{g}_j+ \frac{1}{2}\mathbf{D}_y\mathbf{\Psi}_0 \cdot\mathbf{D}_x\mathbf{T}_0 - \frac{1}{2}\mathbf{D}_x\mathbf{\Psi}_0\cdot\mathbf{D}_y\mathbf{T}_0
\end{align}

For $\mbox{Nu}_2(\epsilon\mathbf{e}_i+\epsilon\mathbf{e}_j)$, $i,j = 1, \ldots, N$, we have ($f_{A}$, $f_{B}$) = ($\mathbf{V}_i+\mathbf{V}_j,\mathbf{0}$) and by linearity of (\ref{T1A})--(\ref{T1B}), ($T_{1A}$, $T_{1B}$) = ($\mathbf{g}_i+\mathbf{g}_j,\mathbf{h}_i+\mathbf{h}_j$). Consequently,
\begin{align}
    \mathbf{b}_{2s}(\epsilon\mathbf{e}_i+\epsilon\mathbf{e}_j) = - \frac{1}{2}\mathbf{D}_y(\mathbf{V}_i+\mathbf{V}_j)\cdot\mathbf{D}_x(\mathbf{g}_i+\mathbf{g}_j) + \frac{1}{2}\mathbf{D}_x(\mathbf{V}_i+\mathbf{V}_j)\cdot\mathbf{D}_y(\mathbf{g}_i+\mathbf{g}_j)+\mathbf{D}_y\mathbf{\Psi}_0 \cdot\mathbf{D}_x\mathbf{T}_0 - \mathbf{D}_x\mathbf{\Psi}_0\cdot\mathbf{D}_y\mathbf{T}_0
\end{align}

We showed in (\ref{rNu}) that Nu$_2$ is a linear function of $\mathbf{b}_{2s}$, i.e. an inner product of $\mathbf{b}_{2s}$ with the constant vector $\mathbf{r}_{\mathrm{Nu}}$. So the left hand side of (\ref{Ablock}) is 
\begin{align}
    \mbox{Nu}_2(\epsilon\mathbf{e}_i\!+\!\epsilon\mathbf{e}_j)-\mbox{Nu}_2(\epsilon\mathbf{e}_i)-\mbox{Nu}_2(\epsilon\mathbf{e}_j) = \mathbf{r}_{\mathrm{Nu}}^T(\mathbf{b}_{2s}(\epsilon\mathbf{e}_i\!+\!\epsilon\mathbf{e}_j)-\mathbf{b}_{2s}(\epsilon\mathbf{e}_i)-\mathbf{b}_{2s}(\epsilon\mathbf{e}_j)). \label{Nu2ABlock}
\end{align}

To verify this equals the right hand side of (\ref{Ablock}), we will show that
\begin{align}
    \mathbf{b}_{2s}(\epsilon\mathbf{e}_i) = \mathbf{b}_{2s}(\epsilon\mathbf{e}_{i+N}) ; \quad \mathbf{b}_{2s}(\epsilon\mathbf{e}_j) = \mathbf{b}_{2s}(\epsilon\mathbf{e}_{j+N}) ; \quad \mathbf{b}_{2s}(\epsilon\mathbf{e}_i\!+\!\epsilon\mathbf{e}_j) = \mathbf{b}_{2s}(\epsilon\mathbf{e}_{i+N}\!+\!\epsilon\mathbf{e}_{j+N}), \quad i,j = 1, \ldots, N \label{b2sABlock}
\end{align}
The terms on the right sides of the equations (\ref{b2sABlock}) correspond to 
perturbations that are sine functions of time. To compute these terms we recall the following property of (\ref{T1A})--(\ref{T1B}) stated in section \ref{sec:Modes}: if the solution for ($T_{1A}$, $T_{1B}$) is ($g,h$) when ($f_{A}$, $f_{B}$) = ($f,0$), then the solution for ($T_{1A}$, $T_{1B}$) is ($-h,g$) when ($f_{A}$, $f_{B}$) = ($0,f$). For the perturbation $\epsilon\mathbf{e}_{i+N}$,  
($f_{A}$, $f_{B}$) = ($\mathbf{0},\mathbf{V}_i$) and therefore ($T_{1A}$, $T_{1B}$) = ($-\mathbf{h}_i,\mathbf{g}_i$), using the result for the perturbation $\epsilon\mathbf{e}_{i}$. Plugging these into the right hand side of equation (\ref{T2s}) for $T_{2s}$ shows that $\mathbf{b}_{2s}(\epsilon\mathbf{e}_i) = \mathbf{b}_{2s}(\epsilon\mathbf{e}_{i+N})$ and likewise for the other two equations in (\ref{b2sABlock}). Thus $\mbox{Nu}_2(\epsilon\mathbf{e}_{i+N}\!+\!\epsilon\mathbf{e}_{j+N})-\mbox{Nu}_2(\epsilon\mathbf{e}_{i+N})-\mbox{Nu}_2(\epsilon\mathbf{e}_{j+N})$ equals the right side of (\ref{Nu2ABlock}), and (\ref{Ablock}) follows.

To show (\ref{Bblock}) we need two terms we have not yet computed,  $\mathbf{b}_{2s}(\epsilon\mathbf{e}_{i}+\epsilon\mathbf{e}_{j+N})$ and 
$\mathbf{b}_{2s}(\epsilon\mathbf{e}_{i+N}+\epsilon\mathbf{e}_j)$.

For the perturbation $\epsilon\mathbf{e}_{i}+\epsilon\mathbf{e}_{j+N}$ we have 
($f_{A}$, $f_{B}$) = ($\mathbf{V}_i,\mathbf{V}_j$). Using the linearity of (\ref{T1A})--(\ref{T1B}) and the results for the two perturbations separately, ($T_{1A}$, $T_{1B}$) = ($\mathbf{g}_i-\mathbf{h}_j,\mathbf{g}_j+\mathbf{h}_i$). Thus 
\begin{align}
    \mathbf{b}_{2s}(\epsilon\mathbf{e}_i+\epsilon\mathbf{e}_{j+N}) =& - \frac{1}{2}\mathbf{D}_y\mathbf{V}_i\cdot\mathbf{D}_x(\mathbf{g}_i-\mathbf{h}_j) - \frac{1}{2}\mathbf{D}_y\mathbf{V}_j\cdot\mathbf{D}_x(\mathbf{g}_j+\mathbf{h}_i) + \frac{1}{2}\mathbf{D}_x\mathbf{V}_i\cdot\mathbf{D}_y(\mathbf{g}_i-\mathbf{h}_j) \nonumber \\&+\frac{1}{2}\mathbf{D}_x\mathbf{V}_j\cdot\mathbf{D}_y(\mathbf{g}_j+\mathbf{h}_i)
    +\mathbf{D}_y\mathbf{\Psi}_0 \cdot\mathbf{D}_x\mathbf{T}_0 - \mathbf{D}_x\mathbf{\Psi}_0\cdot\mathbf{D}_y\mathbf{T}_0 \label{ijN}
\end{align}

For the perturbation $\epsilon\mathbf{e}_{i+N}+\epsilon\mathbf{e}_{j}$ we have 
($f_{A}$, $f_{B}$) = ($\mathbf{V}_i,\mathbf{V}_j$). The result for 
$\mathbf{b}_{2s}$ is the same as (\ref{ijN}) but with $i$ and $j$ switched:
\begin{align}
    \mathbf{b}_{2s}(\epsilon\mathbf{e}_{i+N}+\epsilon\mathbf{e}_{j}) =& - \frac{1}{2}\mathbf{D}_y\mathbf{V}_j\cdot\mathbf{D}_x(\mathbf{g}_j-\mathbf{h}_i) - \frac{1}{2}\mathbf{D}_y\mathbf{V}_i\cdot\mathbf{D}_x(\mathbf{g}_i+\mathbf{h}_j) + \frac{1}{2}\mathbf{D}_x\mathbf{V}_j\cdot\mathbf{D}_y(\mathbf{g}_j-\mathbf{h}_i) \nonumber \\&+\frac{1}{2}\mathbf{D}_x\mathbf{V}_i\cdot\mathbf{D}_y(\mathbf{g}_i+\mathbf{h}_j)
    +\mathbf{D}_y\mathbf{\Psi}_0 \cdot\mathbf{D}_x\mathbf{T}_0 - \mathbf{D}_x\mathbf{\Psi}_0\cdot\mathbf{D}_y\mathbf{T}_0 \label{iNj}
\end{align}
Combining the first two equations of (\ref{b2sABlock}) with (\ref{ijN}) and (\ref{iNj}), we have
\begin{align}
    \mathbf{b}_{2s}(\epsilon\mathbf{e}_i+\epsilon\mathbf{e}_{j+N})+\mathbf{b}_{2s}(\epsilon\mathbf{e}_{i+N}+\epsilon\mathbf{e}_{j})-\mathbf{b}_{2s}(\epsilon\mathbf{e}_i) - \mathbf{b}_{2s}(\epsilon\mathbf{e}_{i+N}) -\mathbf{b}_{2s}(\epsilon\mathbf{e}_j) - \mathbf{b}_{2s}(\epsilon\mathbf{e}_{j+N}) = 0. \label{b2sSum}
\end{align}
Taking the scalar product of the left side of (\ref{b2sSum}) with 
$\mathbf{r}_{\mathrm{Nu}}$ gives (\ref{Bblock}).

\section{The Hessian is block diagonal\label{app:BlockDiagonal}}
We show now that the Hessian is block diagonal, with the blocks consisting of the Hessians for each period alone. Suppose the flow is of the form (\ref{Psi}) but has modes with at least two different periods $\tau_1$ and $\tau_2$. We write out just four such modes, the sine and cosine terms for each period: 

\begin{align}
    \mathbf{\Psi} = \frac{1}{\sqrt{1+\|\mathbf{a}\|^2}}\left( \mathbf{\Psi}_0 + \mathrm{a}_{1} \mathbf{W}_1  \cos\left(\frac{2\pi t}{\tau_1}\right) +  \mathrm{a}_{2} \mathbf{W}_2  \sin\left(\frac{2\pi t}{\tau_1}\right) + \mathrm{a}_{3} \mathbf{W}_3  \cos\left(\frac{2\pi t}{\tau_2}\right) +  \mathrm{a}_{4} \mathbf{W}_4  \sin\left(\frac{2\pi t}{\tau_2}\right) +\ldots \right). \label{psiW}
\end{align}
We now show that
\begin{align}
     D^2\mbox{Nu}_{ij} \bigg|_{\mathbf{a} = \mathbf{0}} 
    &= \mbox{Nu}_2(\epsilon\mathbf{e}_i\!+\!\epsilon\mathbf{e}_j)-\mbox{Nu}_2(\epsilon\mathbf{e}_i)-\mbox{Nu}_2(\epsilon\mathbf{e}_j) = 0\label{NuPeriods}
\end{align}
if $i$ and $j$ correspond to modes with different periods, i.e.~if $i \in \{1,2\}$ and $j \in \{3,4\}$ or vice versa. Assume the former---$i \in \{1,2\}$ and $j \in \{3,4\}$---without loss of generality. 
Next, we write the continuous version of the flow terms given in (\ref{psiW}),
\begin{align}
\psi(x,y,t) = &\frac{1}{\sqrt{1+c_1\epsilon^2}}\psi_0(y) + \frac{\epsilon}{\sqrt{1+c_1\epsilon^2}}f_{A1}(x,y)\cos(2\pi t/\tau_1)
+ \frac{\epsilon}{\sqrt{1+c_1\epsilon^2}}f_{B1}(x,y)\sin(2\pi t/\tau_1)\\
 &+ \frac{\epsilon}{\sqrt{1+c_1\epsilon^2}}f_{A2}(x,y)\cos(2\pi t/\tau_2)
+ \frac{\epsilon}{\sqrt{1+c_1\epsilon^2}}f_{B2}(x,y)\sin(2\pi t/\tau_2)\label{psipertPeriods}
\end{align}
and then use the expansions in section \ref{Model} to calculate the three Nu$_2$ terms in (\ref{NuPeriods}). (\ref{psipertPeriods})
is similar to equation (\ref{psipert}) but with separate terms for each period. For $\mbox{Nu}_2(\epsilon\mathbf{e}_i)$, we have just the mode-$i$ perturbation, so in (\ref{psipertPeriods}) only one of $f_{A1}$ and $f_{B1}$ is nonzero, $f_{A2} = f_{B2} = 0$, and $c_1 = 1$. Following the derivation in section \ref{Model}, the zeroth- and first-order advection diffusion equations (\ref{T0}) and (\ref{T1}) still hold. With the flow (\ref{psipertPeriods}) inserted into (\ref{T1}), we obtain a coupled system like (\ref{T1A}) and (\ref{T1B}) and a solution pair ($T_{1A1}$, $T_{1B1}$) for period $\tau_1$.
We insert 
\begin{align}
\psi_1 &= f_{A1}(x,y)\cos(2\pi t/\tau_1) + f_{B1}(x,y)\sin(2\pi t/\tau_1). \label{psi11}\\
T_1 &= T_{1A1}(x,y)\cos(2\pi t/\tau_1) + T_{1B1}(x,y)\sin(2\pi t/\tau_1). \label{T11}
\end{align}
into the second-order equation (\ref{T2}). Only steady terms on the right hand side will contribute to $T_{2s}$ and therefore to Nu$_2(\epsilon\mathbf{e}_i)$. The steady terms are 
\begin{align}
    - \frac{1}{2}\partial_yf_{A1} \partial_xT_{1A1} + \frac{1}{2}\partial_xf_{A1}\partial_yT_{1A1}
    - \frac{1}{2}\partial_yf_{B1} \partial_xT_{1B1} + \frac{1}{2}\partial_xf_{B1}\partial_yT_{1B1} + \frac{1}{2}\partial_y\psi_0 \partial_xT_0 - \frac{1}{2}\partial_x\psi_0\partial_yT_0. \label{T2s1}
\end{align}
almost the same as the right hand side of equation (\ref{T2s}), with $c_1 = 1$.

For $\mbox{Nu}_2(\epsilon\mathbf{e}_j)$, we have just the mode-$j$ perturbation, so in (\ref{psipertPeriods}) only one of $f_{A2}$ and $f_{B2}$ is nonzero, $f_{A1} = f_{B1} = 0$, and $c_1 = 1$. The calculation proceeds very similarly to previous case. We obtain a coupled system like (\ref{T1A}) and (\ref{T1B}) and a solution pair ($T_{1A2}$, $T_{1B2}$) for period $\tau_2$.
We now insert 
\begin{align}
\psi_1 &= f_{A2}(x,y)\cos(2\pi t/\tau_2) + f_{B2}(x,y)\sin(2\pi t/\tau_2). \label{psi12} \\
T_1 &= T_{1A2}(x,y)\cos(2\pi t/\tau_2) + T_{1B2}(x,y)\sin(2\pi t/\tau_2). \label{T12}
\end{align}
into the second-order equation (\ref{T2}). The steady terms on the right hand side are now 
\begin{align}
    - \frac{1}{2}\partial_yf_{A2} \partial_xT_{1A2} + \frac{1}{2}\partial_xf_{A2}\partial_yT_{1A2}
    - \frac{1}{2}\partial_yf_{B2} \partial_xT_{1B2} + \frac{1}{2}\partial_xf_{B2}\partial_yT_{1B2} + \frac{1}{2}\partial_y\psi_0 \partial_xT_0 - \frac{1}{2}\partial_x\psi_0\partial_yT_0. \label{T2s2}
\end{align}

For $\mbox{Nu}_2(\epsilon\mathbf{e}_i\!+\!\epsilon\mathbf{e}_j)$,
only one of $f_{A1}$ and $f_{B1}$ is nonzero, only one of $f_{A2}$ and $f_{B2}$ is nonzero, and $c_1 = 2$ in (\ref{psipertPeriods}). We now insert
\begin{align}
\psi_1 &= f_{A1}(x,y)\cos(2\pi t/\tau_1) + f_{B1}(x,y)\sin(2\pi t/\tau_1) + f_{A2}(x,y)\cos(2\pi t/\tau_2) + f_{B2}(x,y)\sin(2\pi t/\tau_2) \label{psi112}
\end{align}
into (\ref{T1}). The right hand side of (\ref{T1}) is linear in $\psi_1$, the left hand side is linear in $T_1$, and the boundary conditions are homogeneous. Since (\ref{psi112}) is a sum of (\ref{psi11}) and (\ref{psi12}), the solution to (\ref{T1}) is
\begin{align}
T_1 &= T_{1A1}(x,y)\cos(2\pi t/\tau_1) + T_{1B1}(x,y)\sin(2\pi t/\tau_1)+ T_{1A2}(x,y)\cos(2\pi t/\tau_2) + T_{1B2}(x,y)\sin(2\pi t/\tau_2), \label{T112}
\end{align}
a sum of (\ref{T11}) and (\ref{T12}). We insert (\ref{psi112}) and (\ref{T112})
into the second-order equation (\ref{T2s}). The steady terms on the right hand side are the sum of the steady terms with mode $i$ and mode $j$ separately, (\ref{T2s1}) and (\ref{T2s2}). All the products of derivatives of terms in (\ref{psi112}) and (\ref{T112}) with different periods yield no steady terms, using trigonometric product-to-sum identities. For $\mbox{Nu}_2(\epsilon\mathbf{e}_i\!+\!\epsilon\mathbf{e}_j)$,
the right hand side of (\ref{T2s}) is the sum of the right hand sides for $\mbox{Nu}_2(\epsilon\mathbf{e}_i)$ and $\mbox{Nu}_2(\epsilon\mathbf{e}_j)$. The boundary conditions are homogeneous and therefore $T_{2s}$ for 
$\mbox{Nu}_2(\epsilon\mathbf{e}_i\!+\!\epsilon\mathbf{e}_j)$ is the sum of $T_{2s}$ for $\mbox{Nu}_2(\epsilon\mathbf{e}_i)$ and $\mbox{Nu}_2(\epsilon\mathbf{e}_j)$. Hence $\mbox{Nu}_2(\epsilon\mathbf{e}_i\!+\!\epsilon\mathbf{e}_j)=\mbox{Nu}_2(\epsilon\mathbf{e}_i)+\mbox{Nu}_2(\epsilon\mathbf{e}_j)$, as claimed. 
We have shown that the Hessian is block diagonal, with a single block for each period.

If $\tau_1$ and $\tau_2$ are commensurate periods, the combined flow (\ref{psiW}) is periodic and Nu and the power are still defined as in (\ref{Nu}) and (\ref{Power}), as period averages. If $\tau_1$ and $\tau_2$ are incommensurate periods, the combined flow (\ref{psiW}) is nonperiodic, and we change the period averages to long-time averages, the limits of (\ref{Nu}) and (\ref{Power}) as $\tau \to \infty$. The above calculations showing the Hessian is block diagonal remain valid in this case.

%%%\end{comment}
%\bibliographystyle{unsrt}
%\bibliography{OptimalHeatTransfer}

\begin{thebibliography}{10}

\bibitem{webb2005enhanced}
Ralph~L Webb and NY~Kim.
\newblock {\em Enhanced heat transfer}.
\newblock Taylor and Francis, NY, 2005.

\bibitem{dipprey1963heat}
Duane~F Dipprey and Rolf~H Sabersky.
\newblock {Heat and momentum transfer in smooth and rough tubes at various Prandtl numbers}.
\newblock {\em International Journal of Heat and Mass Transfer}, 6(5):329--353, 1963.

\bibitem{gee1980forced}
Dennis~Leroy Gee and RL~Webb.
\newblock {Forced convection heat transfer in helically rib-roughened tubes}.
\newblock {\em International Journal of Heat and Mass Transfer}, 23(8):1127--1136, 1980.

\bibitem{hart1985heat}
JC~Hart, JS~Park, and CK~Lei.
\newblock {Heat transfer enhancement in channels with turbulence promoters}.
\newblock {\em Journal of Engineering for Gas Turbines and Power}, 107:628--635, 1985.

\bibitem{fiebig1991heat}
Martin Fiebig, Peter Kallweit, Nimai Mitra, and Stefan Tiggelbeck.
\newblock {Heat transfer enhancement and drag by longitudinal vortex generators in channel flow}.
\newblock {\em Experimental Thermal and Fluid Science}, 4(1):103--114, 1991.

\bibitem{tsia1999measurements}
JP~Tsia and JJ~Hwang.
\newblock {Measurements of heat transfer and fluid flow in a rectangular duct with alternate attached--detached rib-arrays}.
\newblock {\em International journal of heat and mass transfer}, 42(11):2071--2083, 1999.

\bibitem{promvonge2010enhanced}
Pongjet Promvonge, Teerapat Chompookham, Sutapat Kwankaomeng, and Chinaruk Thianpong.
\newblock {Enhanced heat transfer in a triangular ribbed channel with longitudinal vortex generators}.
\newblock {\em Energy Conversion and Management}, 51(6):1242--1249, 2010.

\bibitem{Castelloes2010}
F.~V. Castelloes, J.~N.~N. Quaresma, and Cotta~R. M.
\newblock Convective heat transfer enhancement in low reynolds number flows with wavy walls.
\newblock {\em Int. J. Heat Mass Transfer}, 53:2022, 2010.

\bibitem{accikalin2007characterization}
Tolga A{\c{c}}{\i}kal{\i}n, Suresh~V Garimella, Arvind Raman, and James Petroski.
\newblock {Characterization and optimization of the thermal performance of miniature piezoelectric fans}.
\newblock {\em International Journal of Heat and Fluid Flow}, 28(4):806--820, 2007.

\bibitem{bergles2013current}
Arthur~E Bergles and Raj~M Manglik.
\newblock Current progress and new developments in enhanced heat and mass transfer.
\newblock {\em Journal of Enhanced Heat Transfer}, 20(1), 2013.

\bibitem{chaudhari2010heat}
Mangesh Chaudhari, Bhalchandra Puranik, and Amit Agrawal.
\newblock Heat transfer characteristics of synthetic jet impingement cooling.
\newblock {\em International Journal of Heat and Mass Transfer}, 53(5-6):1057--1069, 2010.

\bibitem{park2016enhancement}
Sung~Goon Park, Boyoung Kim, Cheong~Bong Chang, Jaeha Ryu, and Hyung~Jin Sung.
\newblock Enhancement of heat transfer by a self-oscillating inverted flag in a poiseuille channel flow.
\newblock {\em International Journal of Heat and Mass Transfer}, 96:362--370, 2016.

\bibitem{glezer2016enhanced}
Ari Glezer, Rajat Mittal, and Silas Alben.
\newblock Enhanced forced convection heat transfer using small scale vorticity concentrations effected by flow driven, aeroelastically vibrating reeds.
\newblock Technical report, Georgia Institute of Technology Atlanta United States, 2016.

\bibitem{gallegos2017flags}
Ralph Kristoffer~B Gallegos and Rajnish~N Sharma.
\newblock Flags as vortex generators for heat transfer enhancement: Gaps and challenges.
\newblock {\em Renewable and Sustainable Energy Reviews}, 76:950--962, 2017.

\bibitem{lee2017heat}
Jae~Bok Lee, Sung~Goon Park, Boyoung Kim, Jaeha Ryu, and Hyung~Jin Sung.
\newblock Heat transfer enhancement by flexible flags clamped vertically in a poiseuille channel flow.
\newblock {\em international journal of heat and mass transfer}, 107:391--402, 2017.

\bibitem{rips2020heat}
Aaron Rips, Kourosh Shoele, and Rajat Mittal.
\newblock Heat transfer enhancement in laminar flow heat exchangers due to flapping flags.
\newblock {\em Physics of Fluids}, 32(6), 2020.

\bibitem{han1985heat}
JC~Han, JS~Park, and CK~Lei.
\newblock Heat transfer enhancement in channels with turbulence promoters.
\newblock {\em Journal of Engineering for Gas Turbines and Power}, 107(3):628--635, 1985.

\bibitem{karwa2013performance}
Rajendra Karwa, Chandresh Sharma, and Nitin Karwa.
\newblock Performance evaluation criterion at equal pumping power for enhanced performance heat transfer surfaces.
\newblock {\em Journal of Solar Energy}, 2013(1):370823, 2013.

\bibitem{alben2017optimal}
Silas Alben.
\newblock Optimal convection cooling flows in general 2d geometries.
\newblock {\em Journal of Fluid Mechanics}, 814:484--509, 2017.

\bibitem{alben2017improved}
Silas Alben.
\newblock Improved convection cooling in steady channel flows.
\newblock {\em Physical Review Fluids}, 2(10):104501, 2017.

\bibitem{bergman2011fundamentals}
Theodore~L Bergman, Theodore~L Bergman, Frank~P Incropera, David~P Dewitt, and Adrienne~S Lavine.
\newblock {\em Fundamentals of heat and mass transfer}.
\newblock John Wiley \& Sons, 2011.

\bibitem{lienhard2013heat}
John~H Lienhard.
\newblock {\em {A Heat Transfer Textbook}}.
\newblock Courier Corporation, 2013.

\bibitem{shah2014laminar}
Ramesh~K Shah and Alexander~Louis London.
\newblock {\em {Laminar flow forced convection in ducts: a source book for compact heat exchanger analytical data}}.
\newblock Academic press, 2014.

\bibitem{leveque1928laws}
M.A. L{\'e}v{\^e}que.
\newblock {Les lois de la transmission de chaleur par convection}.
\newblock {\em Les Annales des Mines: Memoires}, 12(13):201--299, 1928.

\bibitem{lamb1932hydrodynamics}
Horace Lamb.
\newblock {\em {Hydrodynamics}}.
\newblock Cambridge University Press, 1932.

\bibitem{graetz1882ueber}
L~Graetz.
\newblock {Ueber die w{\"a}rmeleitungsf{\"a}higkeit von fl{\"u}ssigkeiten}.
\newblock {\em Annalen der Physik}, 254(1):79--94, 1882.

\bibitem{alben2023transition}
Silas Alben.
\newblock Transition to branching flows in optimal planar convection.
\newblock {\em Phys. Rev. Fluids}, 8:074502, Jul 2023.

\bibitem{motoki2018maximal}
Shingo Motoki, Genta Kawahara, and Masaki Shimizu.
\newblock Maximal heat transfer between two parallel plates.
\newblock {\em Journal of Fluid Mechanics}, 851:R4, 2018.

\bibitem{souza2020wall}
Andre~N Souza, Ian Tobasco, and Charles~R Doering.
\newblock Wall-to-wall optimal transport in two dimensions.
\newblock {\em Journal of Fluid Mechanics}, 889:A34, 2020.

\bibitem{kumar2022three}
Anuj Kumar.
\newblock Three dimensional branching pipe flows for optimal scalar transport between walls.
\newblock {\em arXiv preprint arXiv:2205.03367}, 2022.

\bibitem{alben2024optimal}
Silas Alben.
\newblock Optimal wall shapes and flows for steady planar convection.
\newblock {\em Journal of Fluid Mechanics}, 984:A43, 2024.

\end{thebibliography}

\end{document}